\documentclass[pra,superscriptaddress,twocolumn,amsmath,amssymb]{revtex4}
\usepackage{amsfonts}
\usepackage{amssymb}
\usepackage{mathtools}
\usepackage{graphicx}
\usepackage{subfigure}
\usepackage{color}
\usepackage[normalem]{ulem}
\usepackage{natbib}
\usepackage{bbold}
\usepackage{placeins}
\usepackage{braket}
\usepackage{soul,xcolor}
\usepackage{epstopdf}
\usepackage{tabu}
\usepackage{array}
\usepackage{upgreek}
\usepackage{multirow}
\usepackage[utf8]{inputenc}
\usepackage[colorlinks = truelinkcolor = blue, urlcolor  = blue, citecolor = blue, anchorcolor = blue]{hyperref}
\hypersetup{
	colorlinks   = true, 
	urlcolor     = blue, 
	linkcolor    = blue, 
	citecolor   = blue 
}

\begin{document}

	\title{Interplay between nonclassicality and $\mathcal{PT}$ symmetry in an effective two level system with open system effects}

	\author{Javid Naikoo}
	\email{naikoo.1@iitj.ac.in}
	\affiliation{Indian Institute of Technology Jodhpur, Jodhpur 342011, India}

	\author{Subhashish Banerjee}
	\email{subhashish@iitj.ac.in}
	\affiliation{Indian Institute of Technology Jodhpur, Jodhpur 342011, India}
	
    \author{Anirban Pathak}
    \email{anirban.pathak@jiit.ac.in}
    \affiliation{Jaypee Institute of Information Technology, A-10, Sector-62, Noida UP-201309, India}
%
%

%
	\begin{abstract}
	A three level atom in $\Lambda$ configuration is reduced to an effective two level system, under appropriate conditions, and its $\mathcal{PT}$ symmetric properties are investigated. This effective qubit system when subjected to a beam-splitter type of interaction, it provides the scope of directly (indirectly) probing the nonclassical properties  of the output (input) state. Here, we study nonclassical properties of the output state by using some well known measures of nonclassical correlations like the measurement induced disturbance,  concurrence and  negativity. The nonclassical features are found to enhance in the $\mathcal{PT}$ symmetric (PTS) phase  compared to  the $\mathcal{PT}$ symmetry broken (PTSB) phase. Further, the output ports of the beam-splitter are subjected to different quantum noise channels, both non-Markovian, e.g., random telegraph noise as well as Markovian, e.g., phase damping, and amplitude damping noise.   The application of noise channels is found to  decrease the degree of nonclassicality, though continuing  to exhibit distinct behavior  in PTS and PTSB phases, with the dominant behavior appearing in the former case.
	\end{abstract}

	\maketitle 
	
	\section{Introduction}
	In  textbook quantum mechanics, one of the fundamental axiom is that the physical observables are represented by the Hermitian operators which always possess  real eigenvalues and conserve the probability \cite{vonNeumannbook}. In particular, the Hamiltonian $H$ generating the time evolution of the system has real eigenvalues and the corresponding time translation operator $U=e^{-i H t}$ is unitary as a consequence of Hermiticity of $H$. However, a non-Hermitian Hamiltonian with the parity ($\mathcal{P}$) - time ($\mathcal{T}$) symmetry, often referred to  as a $\mathcal{PT}$ symmetric (PTS) Hamiltonian, can also possess real eigenvalue spectrum  \cite{PhysRevLett.80.5243}.  Such non-Hermitian Hamiltonians may  undergo a spontaneous transition to $\mathcal{PT}$ symmetry broken (PTSB) phase \cite{PhysRevA.96.043810}.  The operators $\mathcal{P}$ and $\mathcal{T}$ are defined by their action on the dynamical variables $\hat{x}$ (the position operator) and $\hat{p}$ (the momentum operator), such that the \textit{linear} operator $\mathcal{P}$ acts as $\hat{p} \rightarrow - \hat{p}$ and $\hat{x} \rightarrow - \hat{x}$, while  the \textit{antilinear} operator $\mathcal{T}$ acts such that $\hat{p} \rightarrow - \hat{p}$, $\hat{x} \rightarrow \hat{x}$. Further, $\mathcal{T}$ also flips the sign of $i = \sqrt{-1}$, i.e., it transforms  $i \rightarrow - i$, such that the commutation relation  $[\hat{x}, \hat{p}] = i$ is preserved.

	 The $\mathcal{PT}$ symmetric Hamiltonians belong to a more general class of pseudo-Hermitian systems \cite{Mostafazadeh2002}. The eigenfunctions of a system Hamiltonian  in $\mathcal{PT}$ symmetric phase are also the eigenfunctions of the $\mathcal{PT}$ operator, i.e., all eigenfunctions are also $\mathcal{PT}$ symmetric. However, in the PTSB phase, some or all the eigenvalues become complex and not all the eigenfunctions of the Hamiltonain possess $\mathcal{PT}$ symmetry. With these interesting properties, the non-Hermitian quantum mechanics has attracted a lot of attention, leading to the exploration of $\mathcal{PT}$ symmetric systems in different domains  viz., quantum field theories \cite{PhysRevD.70.025001}, open quantum systems \cite{Rotter_2009,banerjee2018open},  the Anderson model for disordered systems \cite{PhysRevB.63.165108},  optical systems with complex refractive indices \cite{Ganainy2007,Makris,Musslimani,Luo2013}, spin models \cite{Li2014,Giorgi2010}. The $\mathcal{PT}$ symmetric systems with spontaneous generation of photons and superradient emission of radiation were also investigated in \cite{Agarwal1,Agarwal2}. 
	 \par 
	 It is worth mentioning here that the non-Hermitian Hamiltonian, in the context of open quantum systems, are often referred to as \textit{effective} Hamiltonians $H_{eff}$, governing the dynamics in a restrictive subspace of the quantum system and appear as von-Neumann type evolution in the Master equations \cite{FESHBACH1958357,Plenio1,Jakob,Rotter2009}.  Thus, the notion of $\mathcal{PT}$ symmetry has proved to be a useful tool in probing the behavior of dynamics of the systems described by effective Hamiltonians which correspond to non-Hermititan systems. Since, the degree of quantumness of a system is controlled by the underlying dynamics,  this naturally invites one to explore the interplay between  nonclassicality and $\mathcal{PT}$ symmetry in such systems \cite{JavidQZE,malpani2019lower,THAPLIYAL2015261,javidprobing,dharcontrollable,AdhikariDiscord,sharma2015}.\par 
	In this work, we will analyze the behavior of   nonclassical correlations, quantified by well known measures of quantum correlations viz., measurement induced disturbance (MID) \cite{PhysRevA.77.022301}, concurrence \cite{chakrabarty2010study,banerjeedynamics,banerjeeentanglement} and negativity, in a $\mathcal{PT}$ symmetric system. This will be achieved by combining the qubit state of the $\mathcal{PT}$ symmetric system with the vacuum at the beam-splitter (or through an interaction which is mathematically equivalent to a beam-splitter operation),  thereby analyzing the resulting output state for the above mentioned nonclassicality measures \cite{Adam2015,AdamIncrease}.\par	
	The paper is organized as follows: In Sec. \ref{NonclasBS}, we discuss beam-splitter operation and how it can be used to  probe the nonclassicality of a single qubit state. This is followed by a discussion of various measures of nonclassicality. Section \ref{Sec:Model} is devoted to detailed discussion of the model consisting of a $\mathcal{PT}$ symmetric system. In Sec. \ref{channeleffect}, we analyze the effect of various quantum noise channels on the nonclassical feature of the output state. Results and their discussion is presented in Sec. \ref{ResultsDiscuss}. We conclude in Sec. \ref{conclusion}.
	\section{Nonclassicality for a single input state at beam-splitter}\label{NonclasBS}
	A single qubit state  when fed to one port of a beam-splitter and  the vacuum at the other port, results in a bipartite state which may exhibit nonclassical properties including entanglement \cite{Asboth05,Adam2015}. Specifically, at the output of the beam-splitter a two-mode entangled state is obtained if and only if the state (other than the vacuum state) fed into the input is a single-mode nonclassical state. Thus, the nonclassicality of the single mode input state gets transferred to a two mode entangled state, and one can try to measure the nonclassicality of the input state by measuring the entanglement of the output state \cite{Asboth05,Adam2015}. For a  particular beam-splitter setting, the behavior of the nonclassical properties of the output state is entirely controlled by the input state parameters. Since, we are not considering optical qubits, in the present study, a beam-splitter operation is visualized as an operation described by an interaction Hamiltonian which is mathematically equivalent to the beam-splitter Hamiltonian.  
	\subsection{Beam splitter input-output state}
    In what follows, we plan to analyze the nonclassicality of the simplest quantum state,  a qubit state, parametrized by $p \in [0,1]$, and  $x$ such that $ |x| \in [0, \sqrt{p(1-p)}]$  and given by 
    \begin{equation}\label{eq:qubit}
	\rho (p; x) \equiv [\rho_{mn}] = \begin{pmatrix}
						1-p  &   x\\
						x^*  &  p
						\end{pmatrix}.
	\end{equation}
    Such states  when combined with the vacuum at  a beam-splitter, result in the output state being separable (if the input qubit state is classical) or entangled (if the input qubit state is nonclassical). One can then use various measures of quantum correlation to directly probe the nonclassicality of the output state and thus indirectly probe the nonclassical properties of the input state. Here the output state can be expressed as
    \begin{equation}\label{BSopt}
    \rho_{out}(\theta) ={\rm U_{BS}} (\rho \otimes |0\rangle \langle 0 |) {\rm U_{BS}}^\dagger,
    \end{equation}
where ${\rm U_{BS}} = {\rm exp}(-\frac{i}{\hbar} H \theta)$  corresponds to a unitary transformation of the beam-splitter operation. The balanced beam-splitter operation is characterized by $\theta = \pi/2$ and can be generated by the Hamiltonian $H = \frac{i \hbar}{2} (a_1^\dagger a_2 - a_1 a_2^\dagger) $, with $a_1 (a_2)$ being the annihilation operators for the two input modes. Physical realization of this Hamiltonian is easy for optical qubits. However, for atomic system, this can be realized by using pulses of appropriate shape and frequency. Specifically, we may note that the balanced beam-splitter operation ${\rm U_{BS}}$ performed on the product input-state $\rho\bigotimes|0\rangle\langle0|$ can be decomposed in terms of standard quantum gates as ${\rm U_{BS}}=({\rm CS})( {\rm T} \bigotimes {\rm T}) \sqrt{{\rm SWAP}}$, where ${\rm CS}$ correspond to a controlled ${\rm S}$ gate. As all these gates can be realized for atomic qubits, an operation equivalent to ${\rm U_{BS}}$ can also be realized for the atomic system of our interest.  In the rest of this work, we will deal with the balanced beam-splitter and call $\rho_{out}(\theta = \pi/2)$ as $\rho_{out}$,  given by
\begin{equation}\label{rhoout}
\rho_{out} = \begin{pmatrix}
                                  1-p                            &  	\frac{i x}{\sqrt{2}}         &	     \frac{x}{\sqrt{2}}    &       	0 \vspace{1mm} \\ 
                                  \frac{-i x^*}{\sqrt{2}}   &      \frac{p}{2}                 &     \frac{-ip}{2}            &        0  \vspace{1mm}\\
                                  \frac{x^*}{\sqrt{2}}      &        \frac{ip}{2}              &        \frac{p}{2}           &       0  \vspace{1mm}\\
                                  0                               &          0                           &           0                     &        0
                              \end{pmatrix}.
\end{equation}
The nonclassicality of this state can be probed using the well known measures such as MID discerning the classical and quantum correlations exhibited by a system under the action of  joint measurements on its subsystems and the entanglement measures such as concurrence and  negativity,  discussed next.\par
\subsection{Measurement induced disturbance}
Consider a bipartite system described by the state $\rho$ belonging to Hilbert space $\mathcal{H}_A \otimes \mathcal{H}_B$, where $\mathcal{H}_{A}$ and $\mathcal{H}_{B}$ represent the state space of systems $A$ and $B$, respectively. One can construct the reduced state for one system by tracing over the other. Let $\rho^{A}$ and $\rho^{B}$ denote the reduced states for $A$ and $B$; one can write
\begin{equation}
\rho^A = \sum\limits_{i} p^A_i \Pi^A_i \quad {\rm and} \quad \rho^B = \sum\limits_{j} p^B_j \Pi^B_j.
\end{equation} 
Here, $\Pi^A$ and $\Pi^B$ are the projectors on the corresponding state space with eigenvalues $p^A$ and $p^B$, respectively. One can define a joint measurement $\Pi$, in terms of the spectral resolution of reduced states, such that the post measurement state is given by
\begin{equation}
\Pi (\rho) = \sum\limits_{i,j}  (\Pi^A_i \otimes \Pi^B_j) \rho (\Pi^A_i \otimes \Pi^B_j).
\end{equation} 
If the post measurement state does not change under the action of $\Pi$, i.e. if $\Pi (\rho) = \rho$, we say that $\rho$ is a classical state with respect to the measurement strategies $\{\Pi^A_i \otimes \Pi^B_j\}$, otherwise $\rho$ is a legitimate quantum state. This idea was used to construct a measure of nonclassicality \cite{PhysRevA.77.022301,PhysRevA.83.064302} given as follows:
\begin{equation}\label{def:MID}
Q(\rho) = I(\rho) - I (\Pi(\rho)).
\end{equation}
Here, $I(\cdot)$ is the quantum mutual information defined as $I(\rho) = S(\rho^A) + S(\rho^B) - S(\rho)$ and $S(\cdot)$ is the von-Neumann entropy. Note that for a classical state $ I (\Pi(\rho)) =  I (\rho)$, so $Q(\rho) = 0$. Therefore, Eq. (\ref{def:MID}) quantifies the difference between the quantum and  classical correlations exhibited by a bipartite system.

\subsection{Entanglement measures} In order to quantify the entanglement in a quantum system, several well known measures have been proposed. These include  entanglement of formation \cite{PhysRevA.53.2046,PhysRevA.56.R3319}, entanglement of distillation\cite{PhysRevA.54.3824}, relative entropy of entanglement\cite{PhysRevLett.78.2275,PhysRevA.57.1619} and negativity  \cite{Eisert,PhysRevA.58.883}. For pure states, the Bell states provide an example of the maximally entangled states. However, in case of mixed states, defining a maximally entangled state is not straightforward.
  In this work, we use two entanglement measures, i.e., concurrence and  negativity. The concurrence, as defined in \cite{PhysRevLett.78.5022}, is given by
  \begin{equation}\label{Conc}
  C(\rho) = {\rm max}\big[ 0, \lambda_1 - \lambda_2- \lambda_3 - \lambda_4\big].
  \end{equation}
  Here, $\lambda_i$ are eigenvalues of the matrix $\sqrt{\sqrt{\rho} \tilde{\rho} \sqrt{\rho}}$ and  $\tilde{\rho} = (\sigma_y \otimes \sigma_y) \rho (\sigma_y \otimes \sigma_y)$  where $\sigma_y$ is the Pauli matrix. Alternatively, $\lambda_i$ represent the square root of the eigenvalues of $\rho \tilde{\rho}$. The parameter $C$ varies between 0 (unentangled states) to  1 (maximally entangled states). The  negativity \cite{Eisert} is based on the positive partial transpose (PPT)  criterion of separability and for  a subsystem $A$ is defined as
  \begin{equation}\label{Neg}
  N(\rho) = \frac{||\rho^{\Gamma_A}||_1 - 1}{2}.
  \end{equation}
  Here, $\rho^{\Gamma_A}$ is the partial transpose of $\rho$ with respect to the subsystem $A$. Equivalently, one can define the \textit{negativity}  as 
  \begin{equation}
  N(\rho) = \sum\limits_{k} \frac{|\lambda_k| - \lambda_k}{2},
  \end{equation}
 where $\{\lambda_k  \}$ is the set of eigenvalues of the partial transposed matrix, $\rho^{\Gamma_A}$. The nonclassical \textit{potential}  for the single mode input state $\rho$ is defined to be the amount of nonclassicality of the output state $\rho_{out}$. Consequently, the concurrence potential (CP) and negativity potential (NP) are defined as \cite{Adam2015} 
  \begin{equation}\label{NCPots}
  CP(\rho) = C(\rho_{out}) \qquad NP(\rho) = N(\rho_{out}).
  \end{equation}
Note that the nonclassical features of the output state are controlled by the input state parameters.\par 
The above mentioned procedure is now applied to a specific system of an effective two level atom interacting with a reservoir. This system exhibits  $\mathcal{PT}$ symmetry in certain parameter range, thereby allowing one to study the interplay between  $\mathcal{PT}$ symmetry and nonclassicality. \par
	
	\section{Model}\label{Sec:Model}
		\begin{figure}[ht] 
		\centering
		\begin{tabular}{cc}
			\includegraphics[width=50mm]{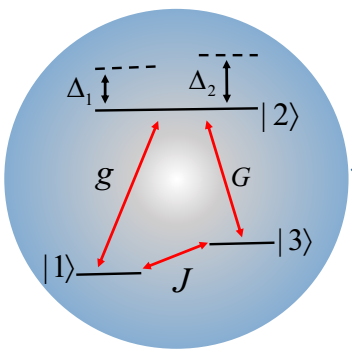}
		\end{tabular}
		\caption{(Color online) Schematic illustration of three-level atom.}
		\label{Model}
	\end{figure}
	 A three level $\Lambda$-type atom, with decay modes associated with all the three levels,  provides an example of a $\mathcal{PT}$ symmetric system \cite{Amarendra,LiOptExpr}. In this work, we will deal with an effective two level Hamiltonian, which is constructed by starting with a $\Lambda$-type system \cite{du2018dynamical} shown in Fig. \ref{Model}.  The resulting qubit state with the general from given by  Eq. (\ref{eq:qubit}), exhibits $\mathcal{PT}$ symmetry and can  be fed to a beam-splitter according to the local operation given in Eq. (\ref{BSopt}). This allows one to explore the nonclassicality of the output bipartite state, Eq. (\ref{rhoout}). This construction, therefore, provides a platform for studying the interplay between  nonclassicality and $\mathcal{PT}$ symmetry of the above described system.
	
	 The Hamiltonian, in a rotating frame with respect to the optical modes, becomes
	\begin{align}
	H &= \hbar \Delta_1 \ket{1}\bra{1} + \hbar \Delta_2 \ket{3}\bra{3} \nonumber \\&- \hbar \bigg[ g\ket{1}\bra{2} + G\ket{3}\bra{2} + \Omega^\prime e^{i\phi} \ket{1}\bra{3} + {\rm H.c.} \bigg].
	\end{align}
	Here, $\Delta_{1,2}$ are the detunings of the optical fields from the corresponding atomic resonances, $g$ and $G$ are Rabi frequencies of the two optical fields. $\Omega^\prime$ is the coupling strength due to the RF-field, the phase $\phi$ can be controlled via the adjustment of the relative phase between RF-field and the two optical fields.
	
	\begin{figure}[h!] 
	\centering
	\includegraphics[width=40mm]{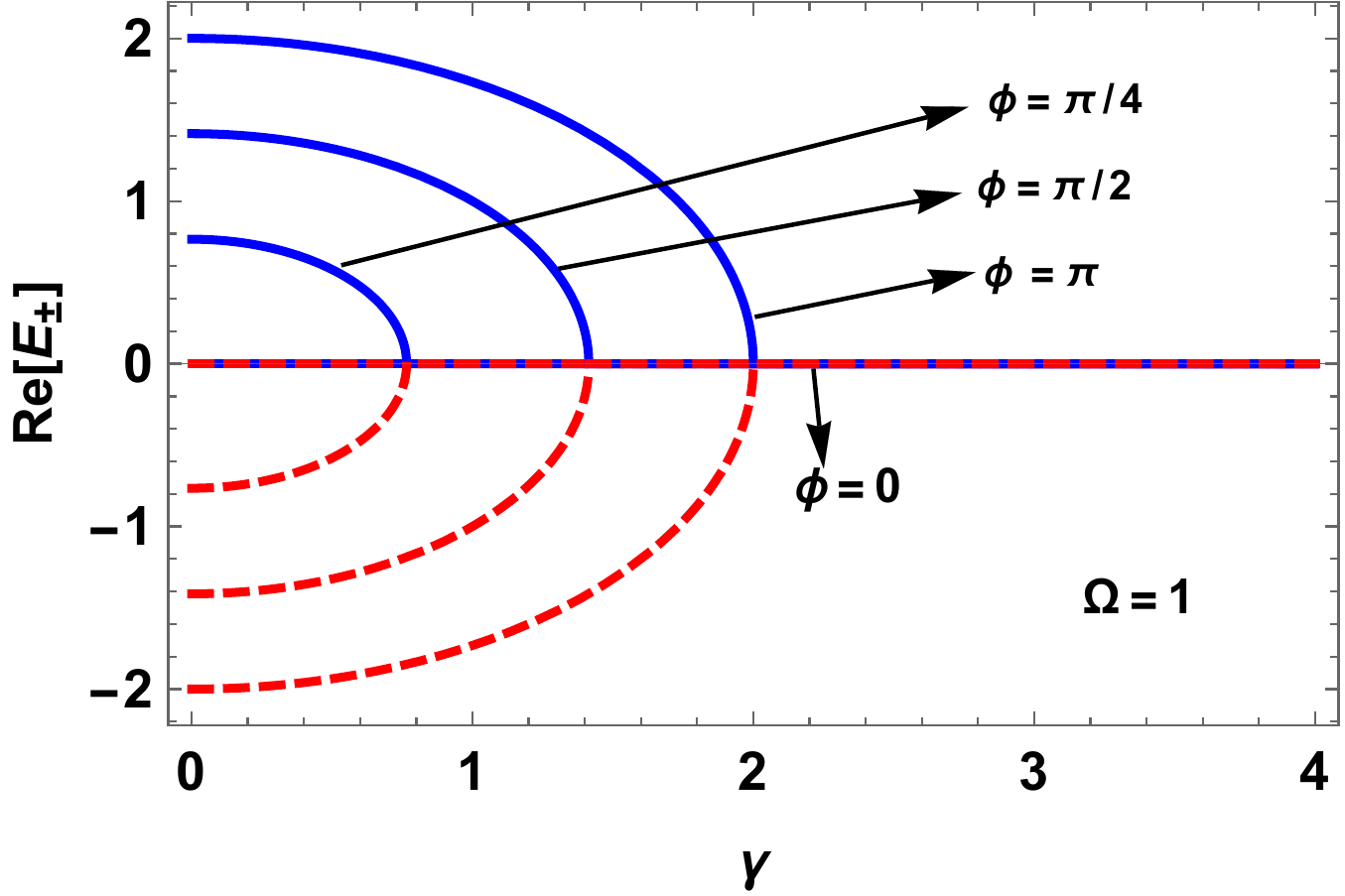}
	\includegraphics[width=40mm]{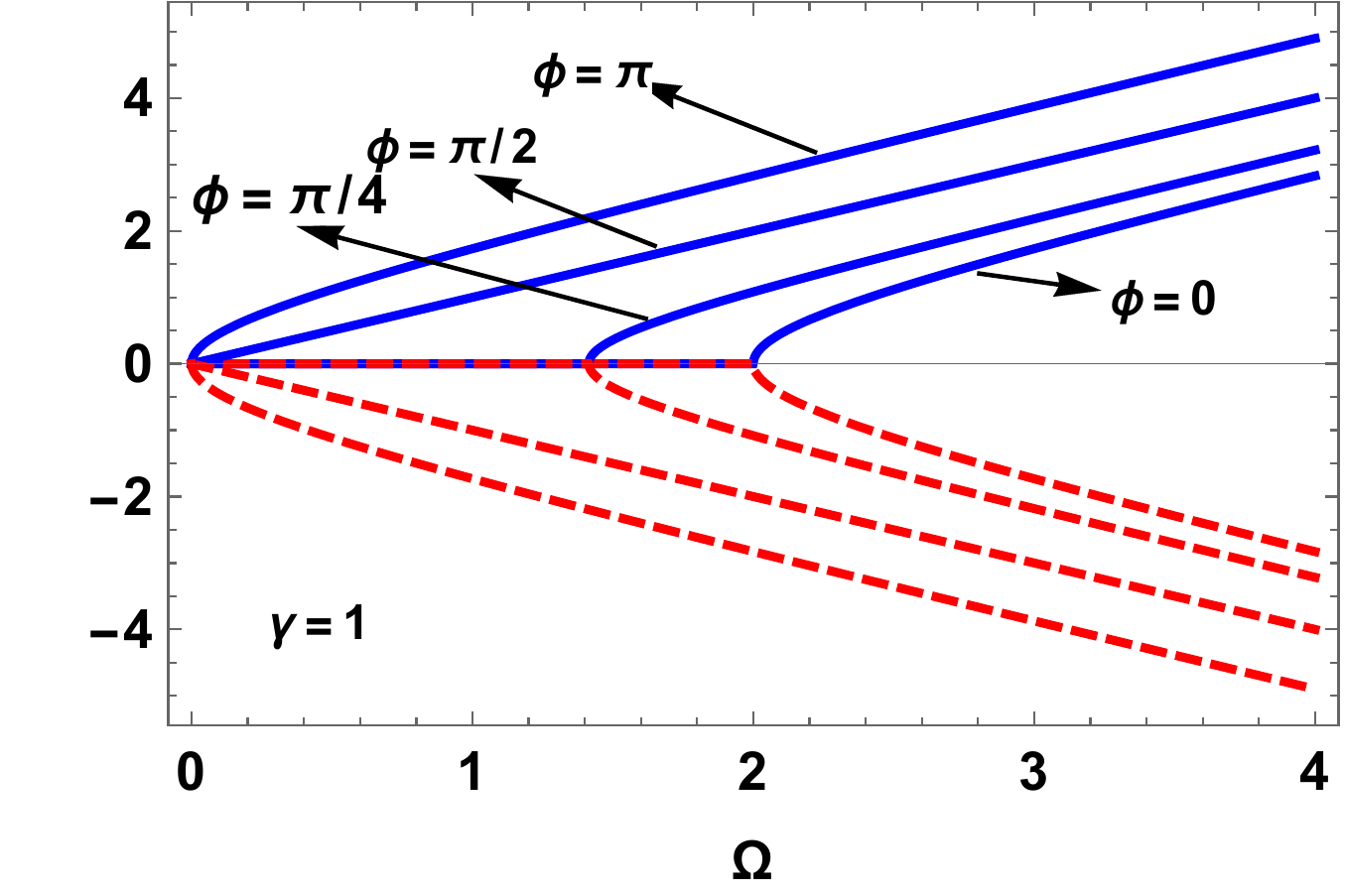}\\
	\includegraphics[width=60mm]{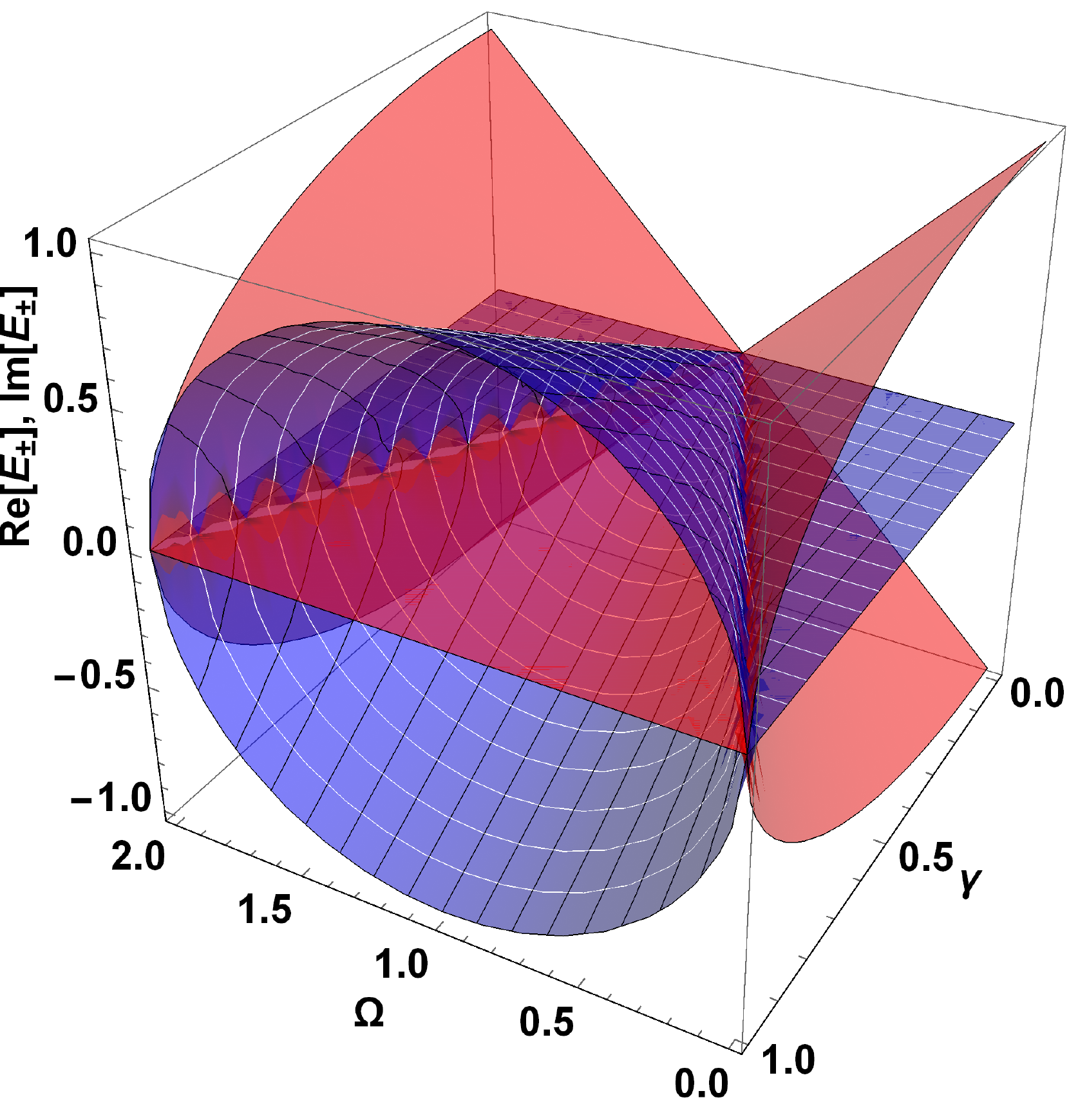}
	\caption{(Color online)  Showing real part of the eigenvalues (top) where blue (solid) and red (dashed) curves correspond to $E_+$ and $E_-$ (Eq. (\ref{eigenvalues})), respectively. Bottom plot shows the real (lined-blue surface) and imaginary (plane-red surface) parts of the eigenvalues.}
	\label{fig:eigenvalues}
\end{figure}
\FloatBarrier

For simplicity, we assume equal coupling strengths  $g=G$ and same detuning $\Delta_1 = \Delta_2 =\Delta$, and use the notation $\frac{\Delta \Omega^\prime}{G^2} = \Omega$. By assuming equal  population gain and loss rates associated with levels $1$ and $3$, respectively,  the following effective two level Hamiltonian can be obtained \cite{du2018dynamical}
	\begin{align}\label{eq:Heff}
	\mathcal{H}_{eff} &= (1- \Omega e^{i \phi}) |1 \rangle \langle 3| + (1- \Omega e^{-i \phi}) |3 \rangle \langle 1| \nonumber \\& + i \gamma |1 \rangle \langle 1| - i \gamma |3 \rangle \langle 3|.
	\end{align}
	The eigenvalues of this Hamiltonian are 
	\begin{equation}\label{eigenvalues}
     E_{\pm} = \pm \sqrt{J^2 - \gamma^2},
	\end{equation}
	with $J = |1 - \Omega e^{i\phi}|$. The eigenvalues are real, and hence the system is $\mathcal{PT}$ symmetric, when the effective coupling $1 - \Omega e^{i\phi}$ is greater than the gain/loss $\gamma$. The $\mathcal{PT}$ symmetry is said to be broken when the gain/loss exceeds the coupling strength. The borderline between the two regimes is such that the eigenvalues $E_{\pm} =0$, called the exceptional points. The variation of energy with respect to the coupling $\Omega$ and gain/loss rate $\gamma$ is shown in Fig. \ref{fig:eigenvalues}.

	Using the effective Hamiltonian of Eq. (\ref{eq:Heff}), one can therefore write the time dependent Schr\"{o}dinger equation  as
	\begin{equation}
	i\frac{\partial \ket{\Psi (t)}}{\partial t} = \mathcal{H}_{eff} \ket{\Psi (t)},
	\end{equation}
	such that the non-unitary time translation operator is given by 
	\begin{equation}
    \mathcal{U}(t) = \begin{pmatrix}
	              \cos(\omega t) + \frac{\gamma}{\omega} \sin(\omega t)                &          -i \frac{1- \Omega e^{i \phi}}{\omega} \sin(\omega t) \\\\
	              -i \frac{1- \Omega e^{-i \phi}}{\omega} \sin(\omega t)               &          \cos(\omega t) - \frac{\gamma}{\omega} \sin(\omega t) 
	              \end{pmatrix}.
	\end{equation}
	Here, $\omega = \sqrt{J^2 - \gamma^2}$. Let  the initial state be $\ket{\Psi (0)} = \ket{1} \equiv (1~~0)^T$ at time $t=0$. Then at some later time $t$, we have
	\small
	\begin{align}
	\ket{\Psi (t)} &= \alpha(t) \ket{1} + \beta(t) \ket{3} \nonumber\\
	               &= \big[ \cos(\omega t) - \frac{\gamma}{\omega} \sin(\omega t) \big] \ket{1}  -i \frac{1- \Omega e^{-i \phi}}{\omega} \sin(\omega t) \ket{3}
	\end{align}
	\normalsize
	such that the probability of the atom being in state $\ket{1}$ and $\ket{3}$ is given by $|\alpha(t)|^2$ and $|\beta(t)|^2$, respectively. We normalize the state vector to have
	\begin{equation}
        \ket{\tilde{\Psi} (t)} = \frac{ \alpha(t) \ket{1} + \beta(t) \ket{3} }{ |\alpha(t)|^2 + |\beta(t)|^2 }.
	\end{equation}
	The corresponding density matrix becomes
	\begin{align}\label{PTqubit}
	\rho(t) &= |\tilde{\Psi}(t) \rangle \langle \tilde{\Psi}(t)| \nonumber\\
	        &= \frac{1}{|\alpha(t)|^2 + |\beta(t)|^2} \begin{pmatrix}
	                                                 |\alpha(t)|^2         &  \alpha(t) \beta^*(t)\\
	                                                 \beta(t) \alpha^*(t)  &   |\beta(t)|^2
	                                                 \end{pmatrix}.
	\end{align}
In what follows, we will use this state as input at the beam-splitter (along with the vacuum state) and analyze the resulting output state for different measures of nonclassicality viz., MID, concurrence and negativity.

 \begin{figure*}[ht] 
	\centering
		\includegraphics[width=55mm]{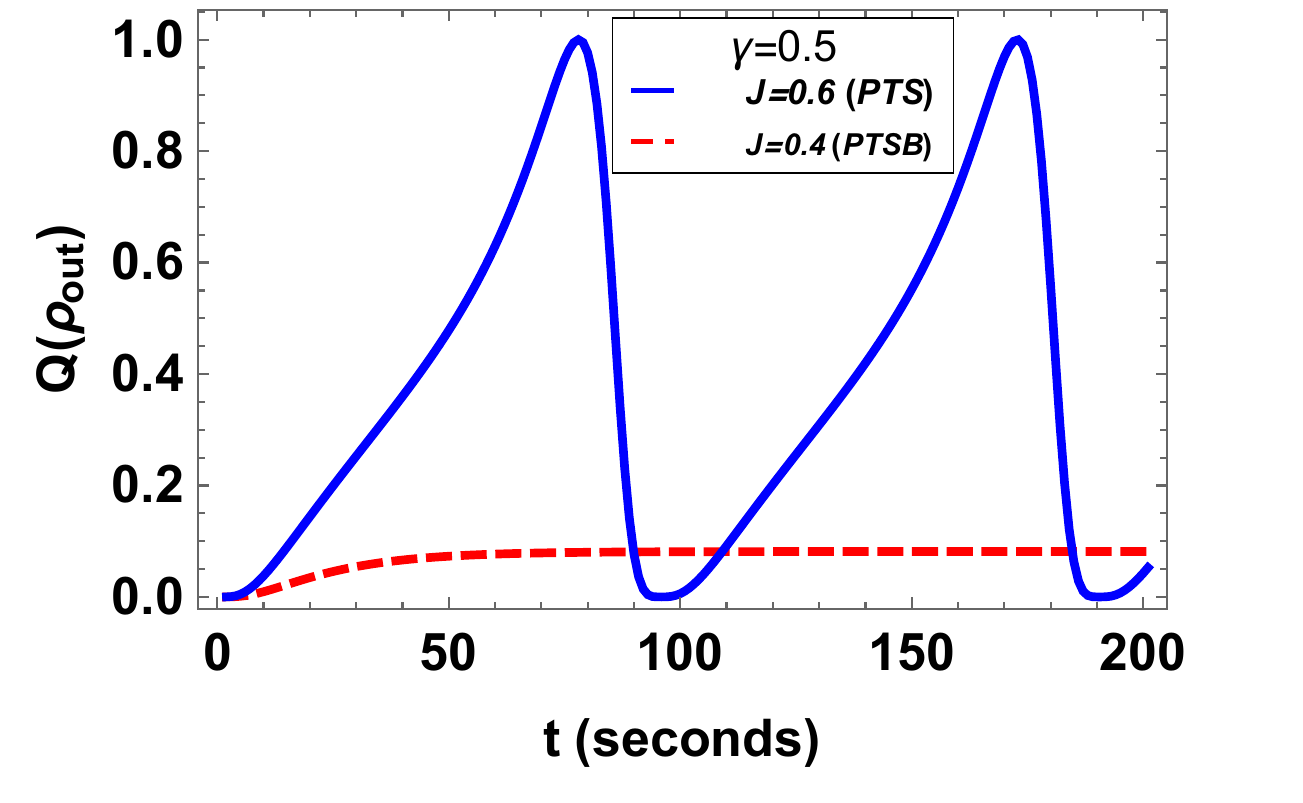}
		\includegraphics[width=59mm]{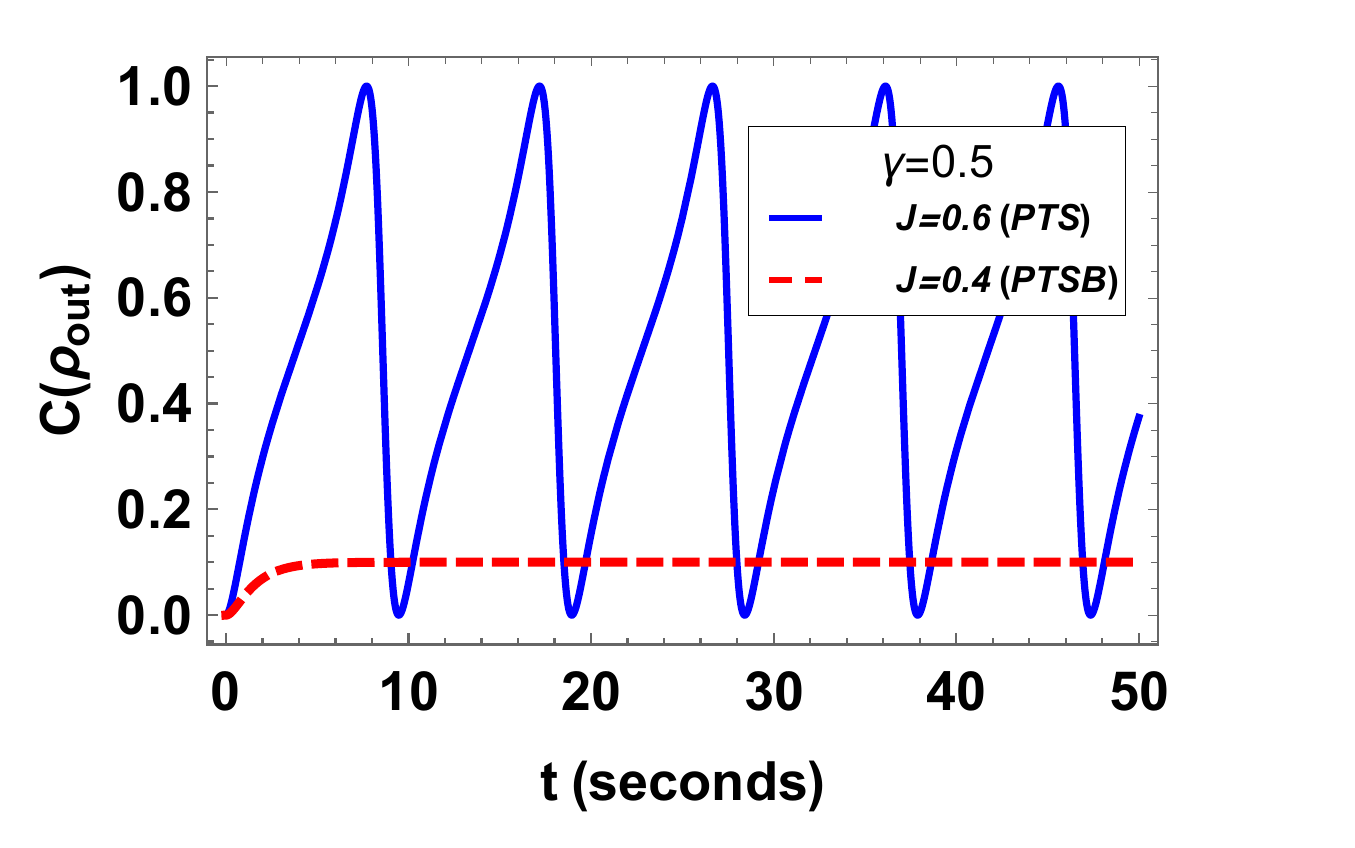}
		\includegraphics[width=59mm]{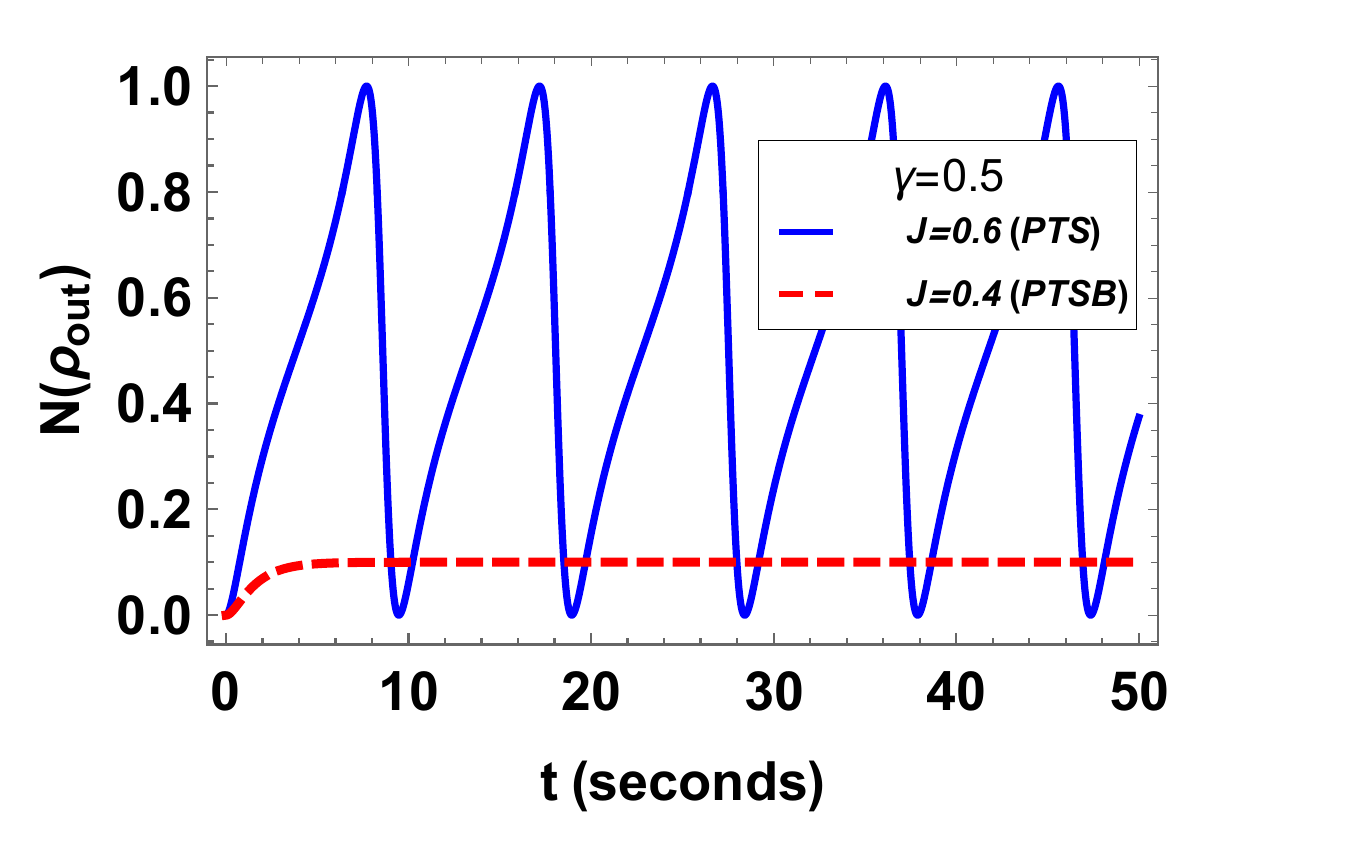}
	\caption{(Color online)  MID  ($Q (\rho_{out})$), Concurrence ($C (\rho_{out})$) and Negativity ($N (\rho_{out})$) of the output state given in Eq. (\ref{rhoout}). The effective coupling between two levels is $J= |1 - \Omega e^{i\phi}|$ and the gain/loss rate is $\gamma$. The conditions  $J>\gamma$ and $J< \gamma$ correspond to PTS and PTSB regimes, respectively. The non-classical features are enhanced in PTS regime, that is, when system coupling dominates the gain/loss rate. This is in consonance with the results of our previous work \cite{QZEjavid}.}
	\label{fig:MID_Con_Neg}
\end{figure*}

\begin{figure}
	\includegraphics[width=75mm]{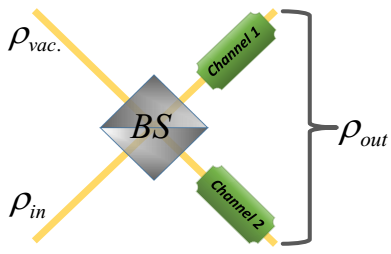}
	\caption{(Color online) Beam-splitter ($BS$) with  input states $\rho_{in}$  and $\rho_{vac} ( = | 0 \rangle \langle 0 |)$. The output ports are subjected to a \textit{channel}, leading to the final output state $\rho_{out}$. The channel parameters are, in general different, unless stated otherwise.}
	\label{fig:BSchannel}
\end{figure}
\section{Effect of different quantum noise channels on nonclassicality}\label{channeleffect}
Evolution of quantum correlations in the presence of non-Markovian noise has been the subject matter of many studies \cite{naikoo2019facets,NMDShrik,kumar2018enhanced,pradeep2,GThomas,thapliyal2017quantum}.  However, to the best of our knowledge, the evolution of the dynamics of the quantum correlations present in a $\mathcal{PT}$ symmetric system has not been investigated earlier in the presence of noise.\par
In this section, we study the interplay between nonclassicality and $\mathcal{PT}$ symmetry when the output ports of the beam-splitter are subjected to different  \textit{quantum noise channels}. Specifically, we  consider Random telegraph noise (RTN) \cite{pradeep1,pradeep2,NMDShrik,naikoo2019facets}, phase damping (PD) \cite{banerjeedynamicsdecoWOdis} and amplitude damping (AD) \cite{SGAD,Banerjee2008} channels. The RTN allows us to bring out the interplay between $\mathcal{PT}$ symmetry and non-Markovian dynamics. Non-Markovian evolution has been found to favor the suppression of decoherence and disentanglement \cite{Amrit2011,Franco2012}. Since one of the major challenges in carrying out  quantum information tasks is to sustain the coherence and entanglement \cite{SBcontrol}, non-Markovianity assisted control on the degree of coherence and entanglement can become very pertinent in future.\par
\textit{Random Telegraph Noise:} This model describes  a qubit subjected to a classical source of random telegraph noise, i.e., a bistable fluctuator randomly switching between its two states with a given rate $\gamma$ \cite{RTNrelated}. The ratio between the switching rate and the system-environment coupling, determines whether the system is Markovian or non-Markovian. The evolution is governed by the following Kraus operators 
\begin{align}\label{KrausRTN}
K_0 (t) &= \sqrt{\frac{1 + \Lambda(t)}{2}}  \begin{pmatrix} 
                                                         1   &   0\\
                                                         0   &   1
                                                      \end{pmatrix}\nonumber \\                                                                 
K_1(t) &= \sqrt{\frac{1 - \Lambda(t)}{2}} \begin{pmatrix}
                                                             1     &    0\\
                                                             0     &    -1              
                                                           \end{pmatrix}.
\end{align}
Here, the parameter $\Lambda(t) = \exp{(- \gamma t) } (\cos( \mu \gamma t )  + \frac{\sin( \mu \gamma t )}{\mu}) $  is called the \textit{memory kernel} and is crucial for determining whether the dynamics is Markovian or non-Markovian. Also, $\mu = \sqrt{(\frac{2 a}{\gamma})^2 - 1}$ and $\gamma = \frac{1}{2 \tau}$. The parameter $a$ is proportional to the strength of system-environment coupling. The Markovian and non-Markovian regimes are characterized by $4 a \tau < 1$ and $4 a \tau > 1$, respectively.
For a general qubit at time $t=0$, given by Eq. (\ref{eq:qubit}), the action of RTN map results in the  state at time $t>0$, give by
\begin{equation}
\rho(t) = \mathcal{E}^{RTN}_{t\leftarrow t_0}[\rho(0)] = \begin{pmatrix}
1-p   &  x \Lambda(t)   \\
x^* \Lambda(t)    &  p
\end{pmatrix}.
\end{equation}
\textit{Phase Damping:}  This noise process is uniquely quantum mechanical in nature and describes the loss of quantum information without loss of energy \cite{banerjee2007dynamics}. In this case, the Kraus operators are given by
\begin{equation}\label{KrausPD}
K_0 = \begin{pmatrix}
           1     &    0\\
           0     & \sqrt{1-\lambda}
          \end{pmatrix}                      \quad {\rm and}\quad     K_1   = \begin{pmatrix}
                                                                                     0     &     0\\
                                                                                     0     &    \sqrt{\lambda}
                                                                                \end{pmatrix}.
\end{equation}
The parameter $\lambda$ can be modeled as $1 - \cos^2(\eta t)$, with $0 \le  \eta t \le \pi/2 $. The action of PD channel on a general state  (Eq. (\ref{eq:qubit})) is as follows

\begin{equation}
\mathcal{E}^{PD}_{t\leftarrow t_0}[\rho] = \begin{pmatrix}
                                                        1-p                          &  x \sqrt{1 - \lambda}   \\
                                                    x^* \sqrt{1 - \lambda}     &  p
                                                      \end{pmatrix}.
\end{equation}

\textit{Amplitude Damping:} This noise process is a schematic model for describing the energy dissipation effects due to loss of energy from a quantum system to its environment. The dynamics is given by the following Kraus operators \cite{srikanth2008squeezed}:
   \begin{equation}\label{KrausAD}
   K_0 = \begin{pmatrix}
                 1     &    0\\
                 0     & \sqrt{1-\gamma}
             \end{pmatrix}                           \quad {\rm and}\quad     K_1   = \begin{pmatrix}
                                                                                                                 0     &     \sqrt{ \gamma}\\
                                                                                                                 0     &    0
                                                                                                            \end{pmatrix},
   \end{equation} 
   The time dependent parameter can be modeled by $\gamma = 1 -  e^{-\chi t}$. Under AD channel a general qubit state evolves as
\begin{equation}
\mathcal{E}^{AD}_{t\leftarrow t_0}[\rho] = \begin{pmatrix}
                                                            1-p (1-\gamma)         &  x \sqrt{1 - \gamma}   \\
                                                        x^* \sqrt{1 - \gamma}     &   p (1-\gamma)
                                                     \end{pmatrix}.
\end{equation}\par
The implementation of the quantum noise channel at the output ports of the beam-splitter can be realized by the combined action of such channels on a bipartite state $\rho(\alpha)$ \cite{AdamHorst}. Note that the channel acts on the output state of the form given by Eq. (\ref{rhoout}) and the input state is given by Eq. (\ref{PTqubit}).
\begin{equation}
\rho(\alpha, q_1, q_2) = \sum\limits_{i, j} \big[ K_i(q_1) \otimes K_j(q_2)\big] \rho(\alpha) \big[ K_i^\dagger(q_1) \otimes K_j^\dagger(q_2)\big]. 
\end{equation}
Here, $q_1$ and $q_2$ are the channel parameter and, in general, $q_1 \neq q_2$, Fig. \ref{fig:BSchannel}. Using this description and the definition in Eq. (\ref{Conc}), one obtains the following analytic expressions for concurrence 
\begin{equation}\label{ConcExps}
C(\rho_{out}) = \begin{cases}
p &  {\rm Noiseless}\\\\
p \Lambda_1 \Lambda_2 &  {\rm RTN}\\\\
p \sqrt{(1-\lambda_1)(1-\lambda_2)} &   {\rm PD}\\\\
p \sqrt{(1-\gamma_1)(1-\gamma_2)} &   {\rm AD}
\end{cases} 
\end{equation}
Here, $p$ is the  probability as defined in Eq. (\ref{eq:qubit}). Unfortunately, the expressions for negativity turn out to be too complicated and are not given here.  An interesting observation is that for $p=0$ (and hence $x=0$), that is, when both  input ports of the beam-splitter contain  vacuum state, the concurrence of the output state becomes zero. The same is true for other nonclassical measures like MID and negativity. This is expected as  the output state is expected to be nonclassical if one of the input states is vacuum and the other one is a nonclassical state, but vacuum state is classical in the sense that it can be described by positive $P$-function as it can be described as a coherent state having 0 photons. The nonclassical measure of the output state in all other cases reflects the degree of nonclassiclaity of the input state.\par 
Another way of looking at the nonclassicality of the post channel output state, is by noting that a two qubit state can be written, as discussed in the Appendix, in the following form:
\begin{equation}\label{psialpha}
\ket{\psi_\alpha} = \sqrt{\alpha} | 0 1\rangle + \sqrt{1 - \alpha} | 10 \rangle.
\end{equation}
This state, when subjected to PD channel, results in the following mixed state
\begin{align*}
\rho_{PD} (q, \lambda_1, \lambda_2) &= (\frac{1}{2} - y) | \beta_1 \rangle \langle \beta_1 |  + (\frac{1}{2} + y) | \beta_2 \rangle \langle \beta_2 | \nonumber\\&+ (\alpha - \frac{1}{2}) (| \beta_1 \rangle \langle \beta_2 | + | \beta_2 \rangle \langle \beta_1 |),
\end{align*}
where $y = \sqrt{\alpha (1-\alpha) (1-\lambda_1)(1-\lambda_2)}$. If one sets $\alpha =\frac{1}{2}$ which means $p=1$, then above state becomes
\begin{align*}
\rho_{PD} (q, \lambda_1, \lambda_2) &= l_+ | \beta_1 \rangle \langle \beta_1 |  + l_- | \beta_2 \rangle \langle \beta_2 |.
\end{align*}
Here, $l_{\pm} = (1 \pm  \sqrt{(1-\lambda_1)(1-\lambda_2)})/2$. Thus, for the special case $p=1$, we have the Bell-diagonal representation of the state Eq. (\ref{psialpha}). The concurrence for such states is given by 
\begin{equation}
C = 2  {\rm max} \big[0, {\rm max}_{\pm}[ l_{\pm}] - \frac{1}{2}\big] = \sqrt{(1-\lambda_1) (1-\lambda_2)}.
\end{equation}
This is  consistent with Eq. (\ref{ConcExps}) for $p=1$.

\begin{figure*}
	\centering
	\includegraphics[width=55mm]{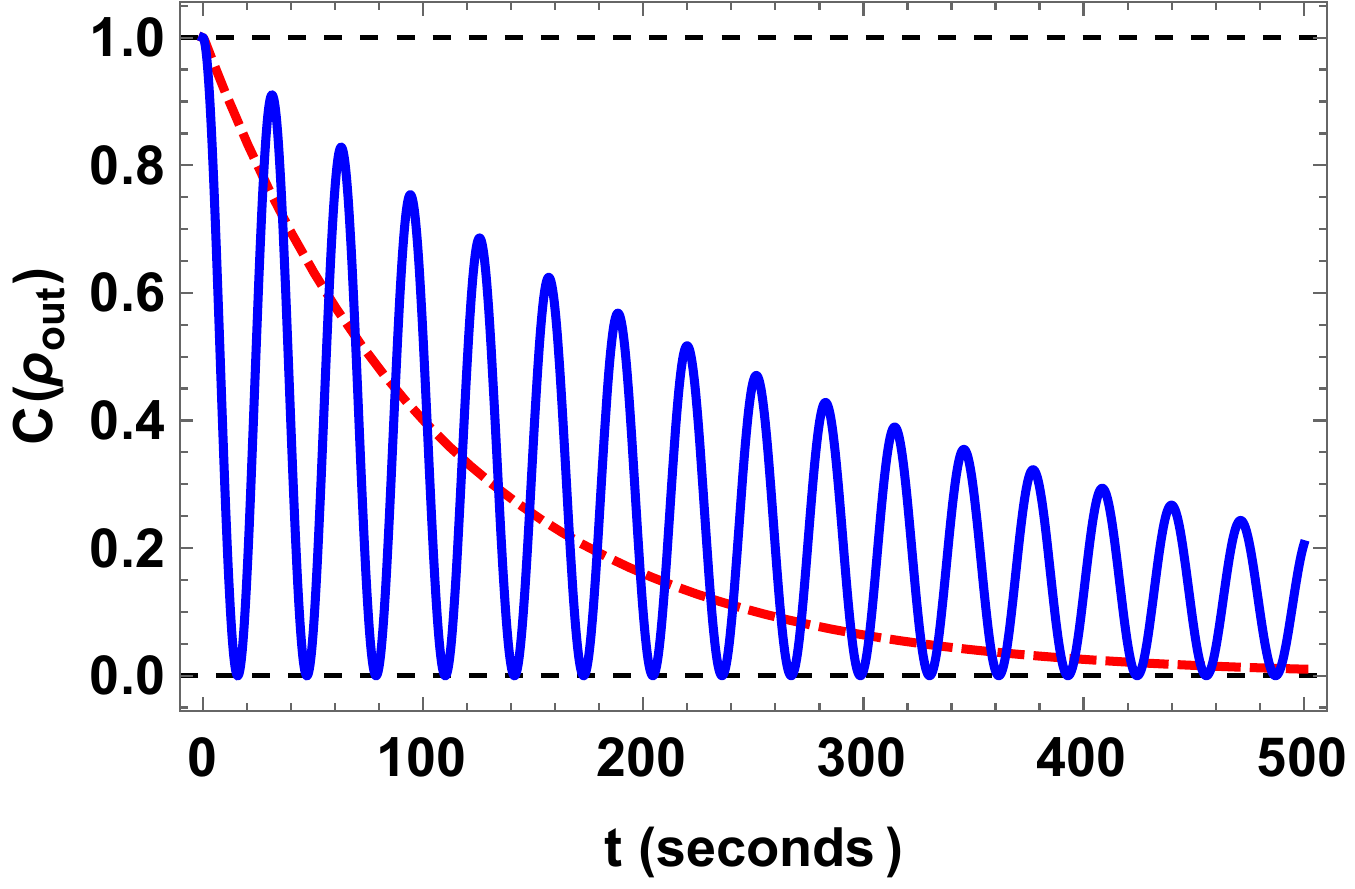}
	\includegraphics[width=55mm]{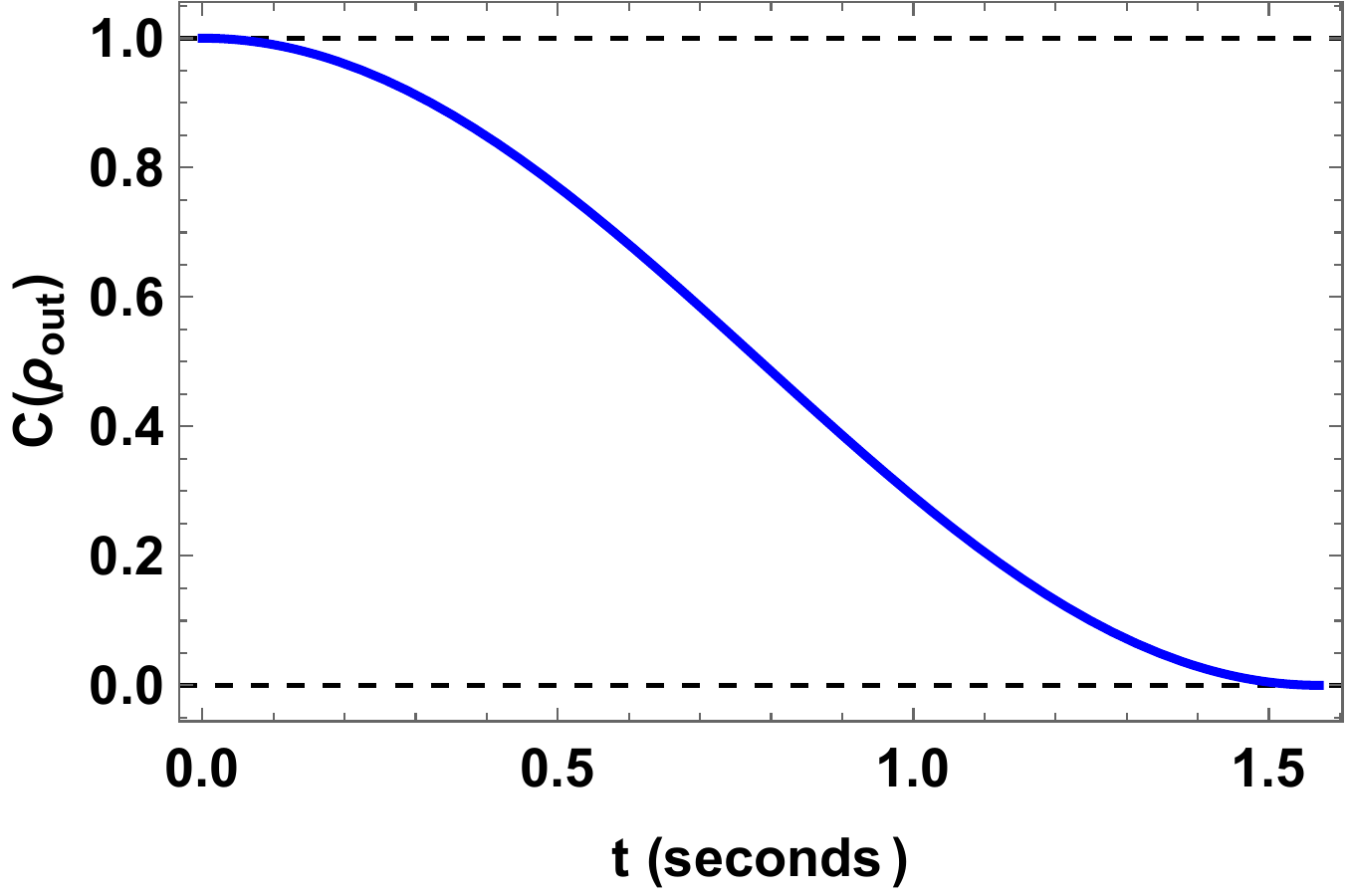}
	\includegraphics[width=55mm]{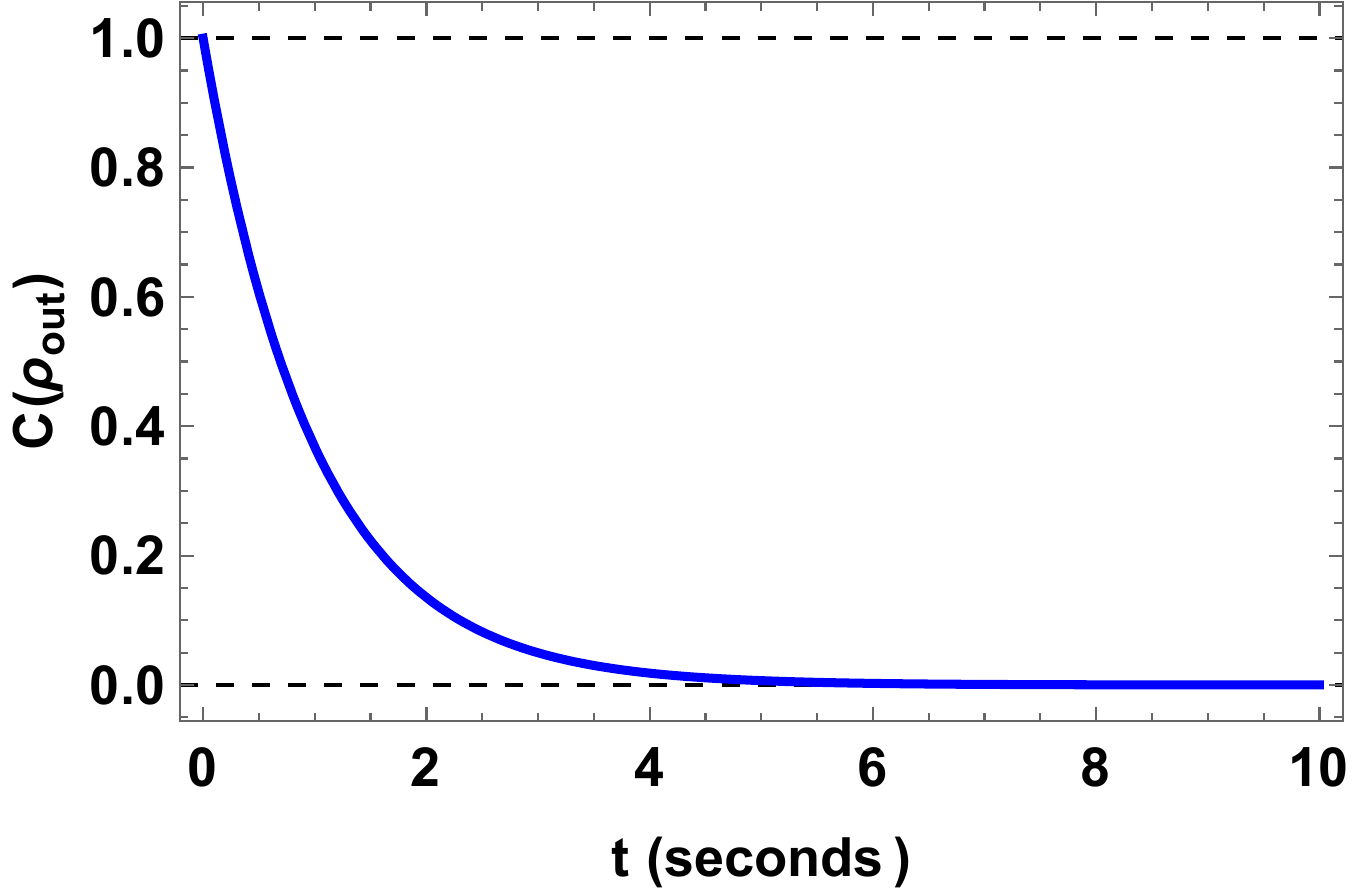}
	\caption{(Color online) Showing  concurrence with respect to time when the state parameter $p=1$ which corresponds to the input state $\ket{1}$ at the beam-splitter. Left, middle and right plots correspond to RTN, PD and AD channels, respectively. The solid (blue) and dashed (red) curves in the left plot pertain to non-Markovian and Markovian processes, respectively.}
	\label{conc_peq1}
\end{figure*}

\begin{figure*}
	\includegraphics[width=65mm]{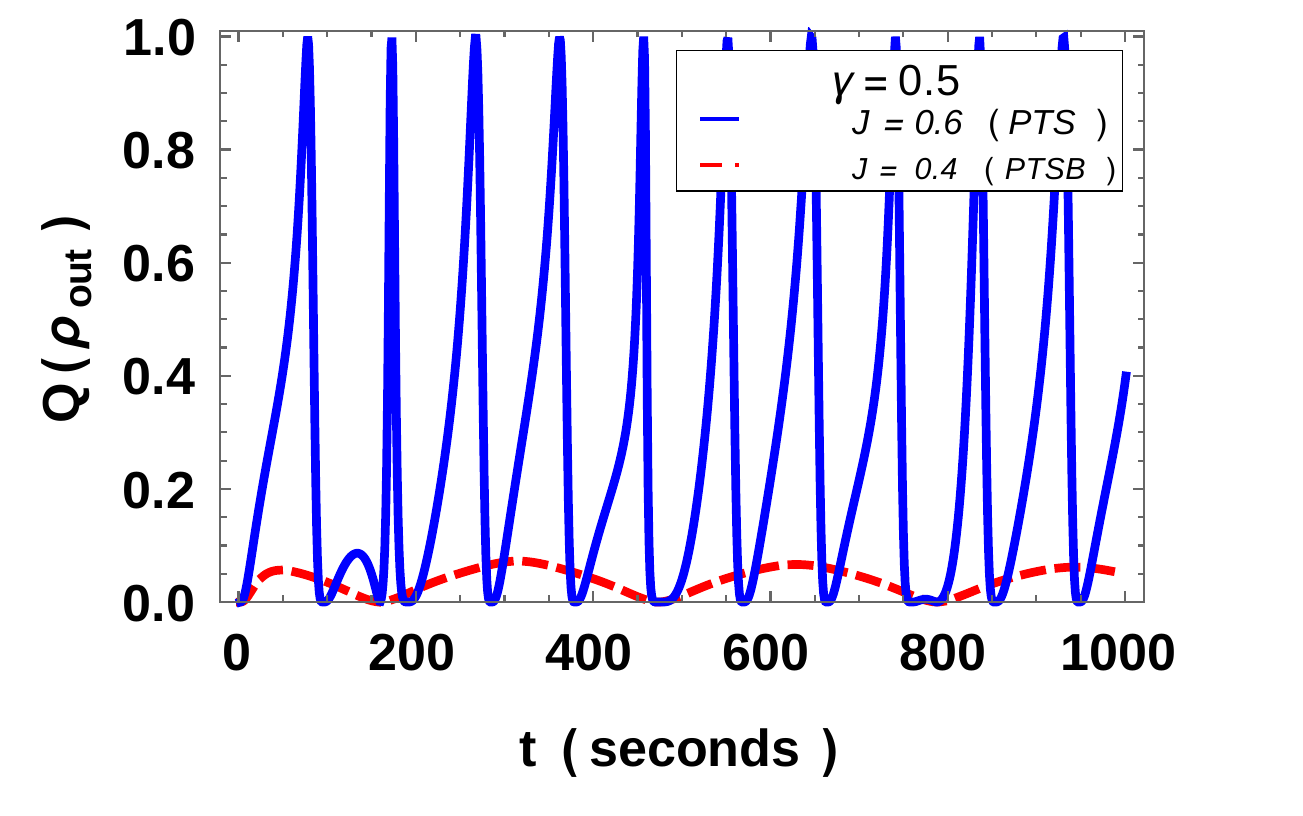}
	\includegraphics[width=65mm]{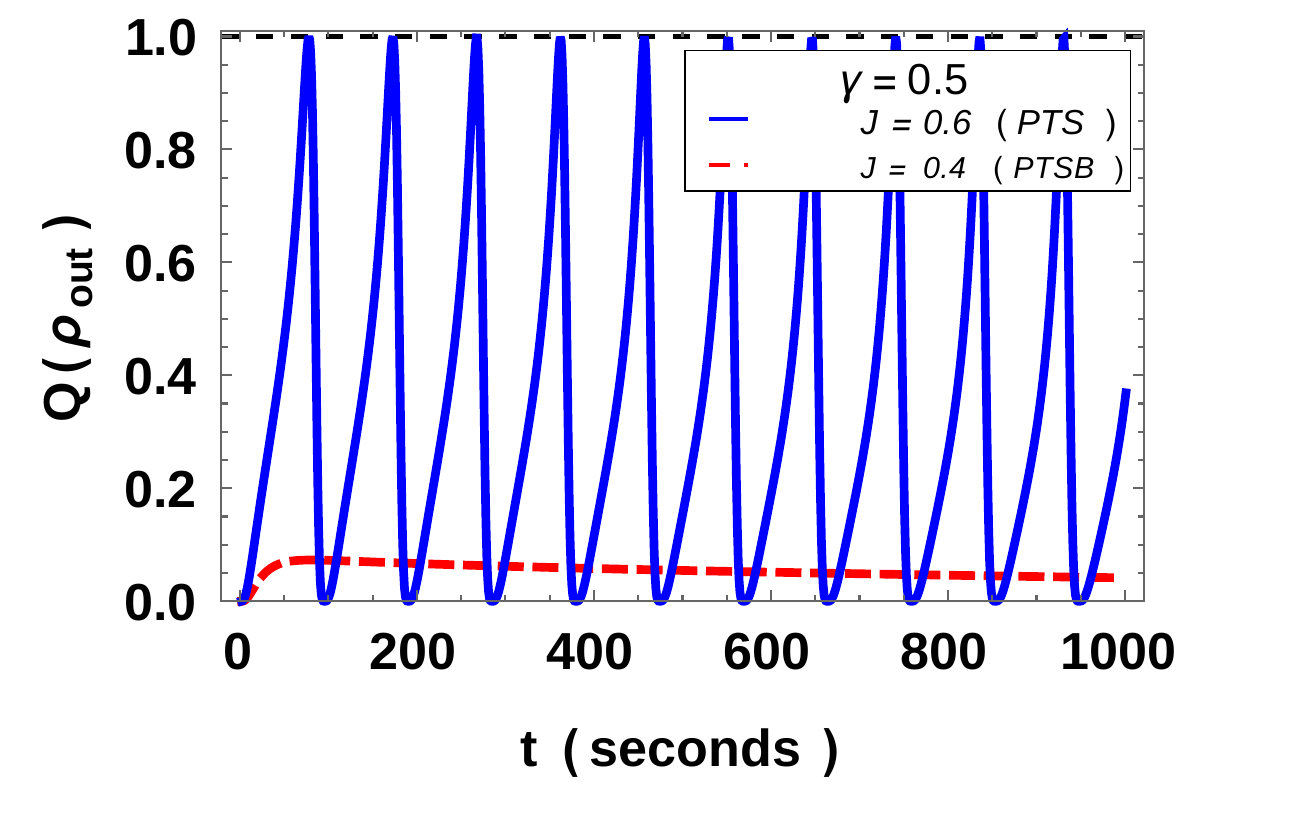}
	\includegraphics[width=65mm]{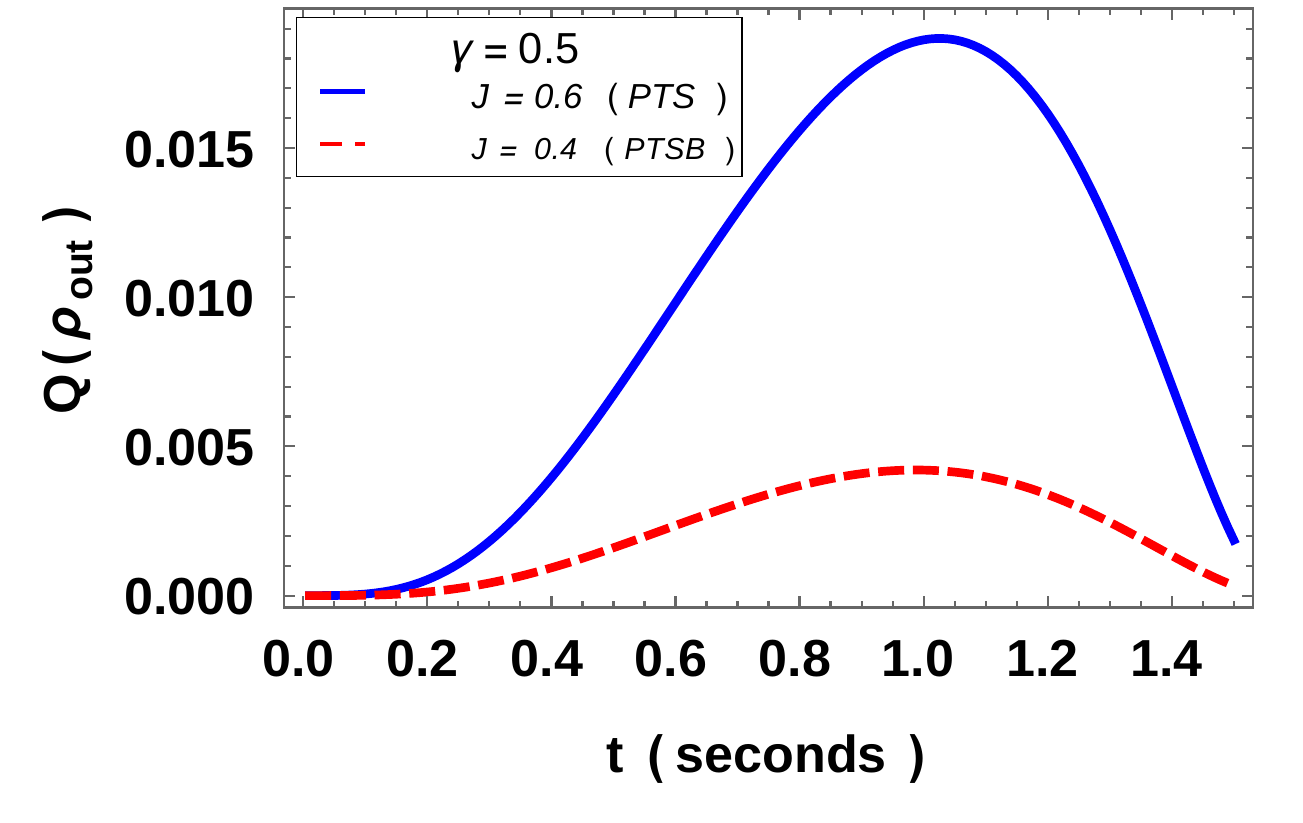}
	\includegraphics[width=65mm]{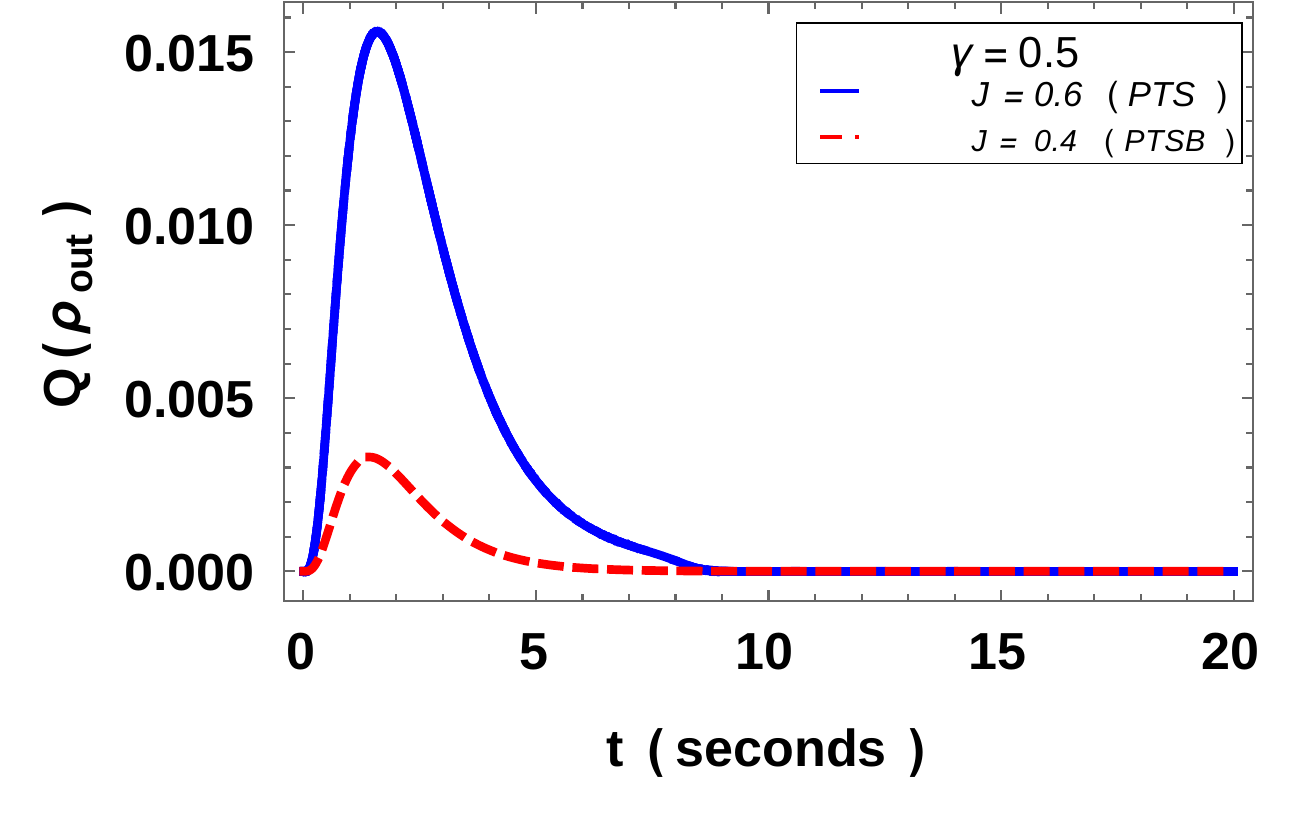}
	
	\caption{(Color online)  MID ($Q (\rho_{out})$), as defined in Eq. (\ref{def:MID}),, is plotted as a function of time $t$ when the output ports of the beam-splitter are subjected to RTN noise in  non-Markovian (top-left) and  Markovina (top-right) regimes, respectively. Same quantity is shown for PD channel (bottom-left) and AD channel (bottom-right). The input state at the beam-splitter is given in Eq. (\ref{PTqubit}).}
	\label{fig:MID}
\end{figure*}

\begin{figure*}
	\includegraphics[width=65mm]{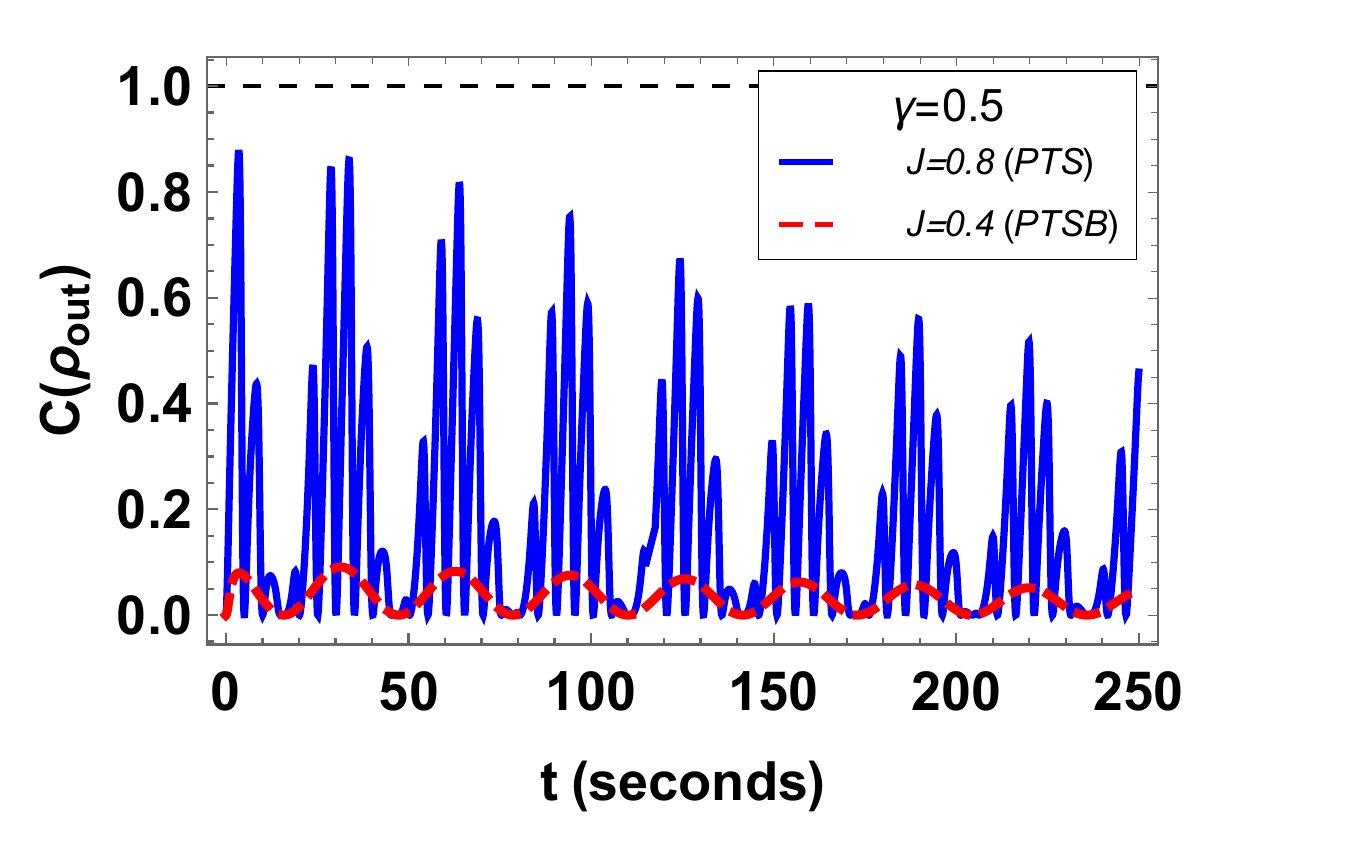}
	\includegraphics[width=65mm]{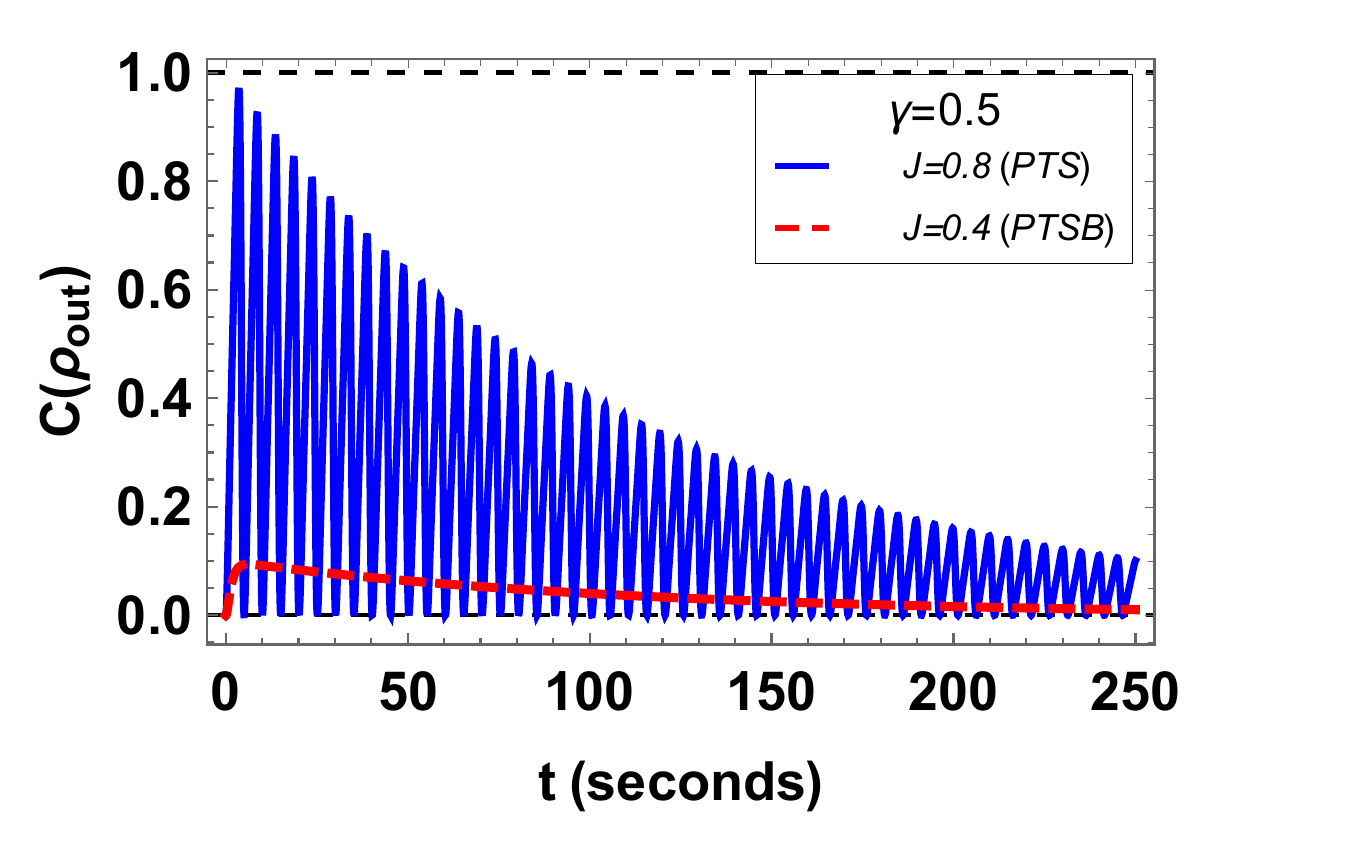}
		\includegraphics[width=65mm]{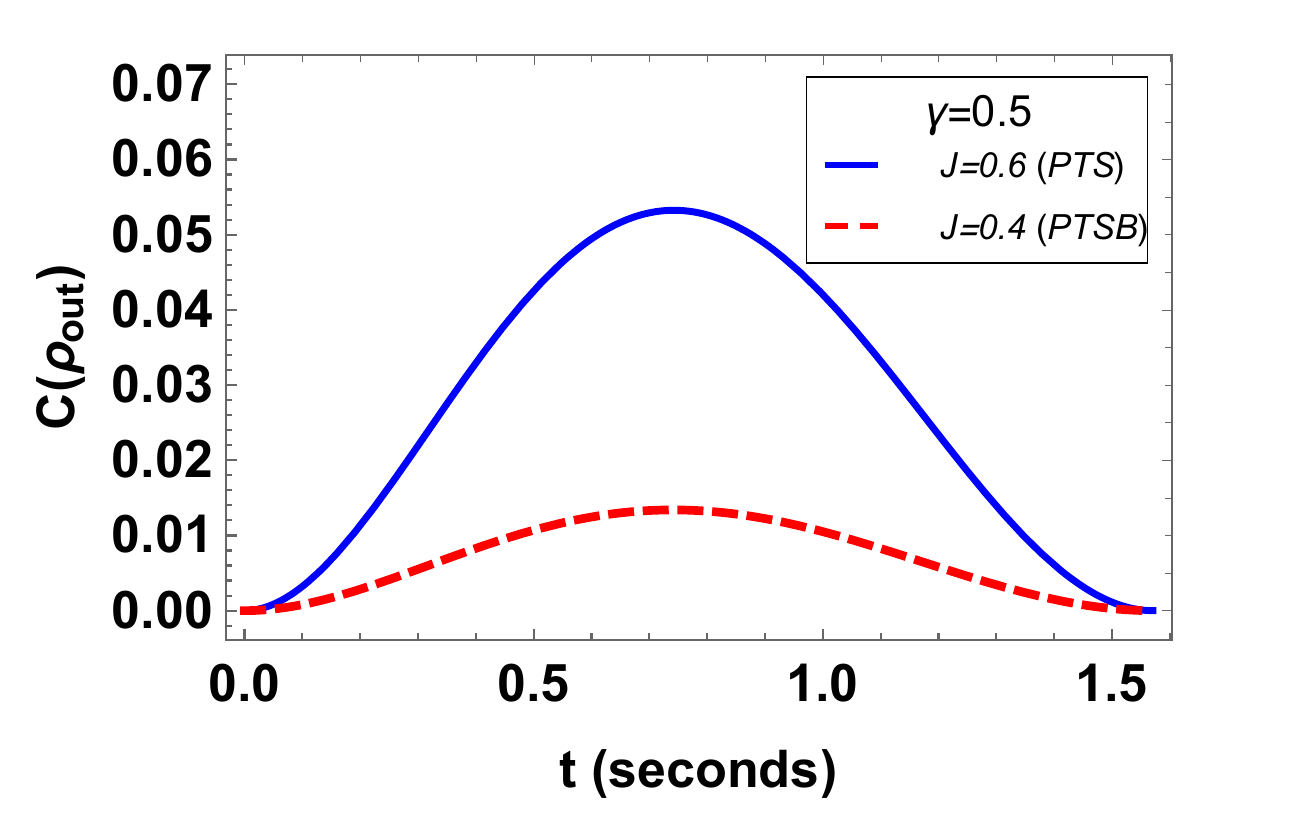}
	\includegraphics[width=65mm]{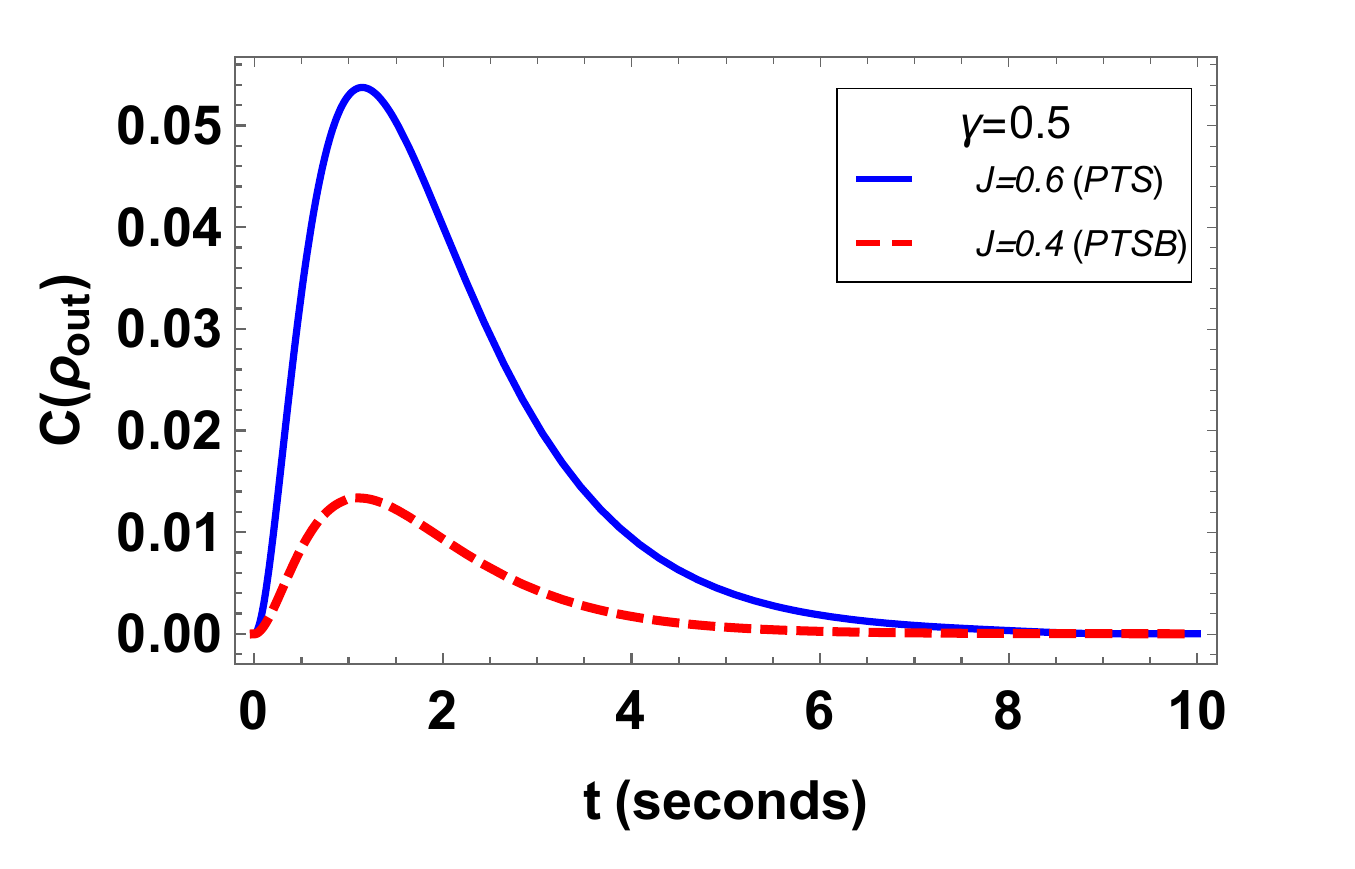}
	
	\caption{(Color online) Concurrence  as a function of time $t$ when the output ports of the beam-splitter are subjected to RTN noise in non-Markovian (top-left) and  Markovina (top-right) regimes, respectively. Same quantity is shown for PD channel (bottom-left) and AD channel (bottom-right). The PTS and PTSB phases are depicted by solid (blue) and dashed (red) curves, respectively. The input state at the beam-splitter is given in Eq. (\ref{PTqubit}).}
	\label{fig:concurrence}
\end{figure*}

\begin{figure}
	\includegraphics[width=65mm]{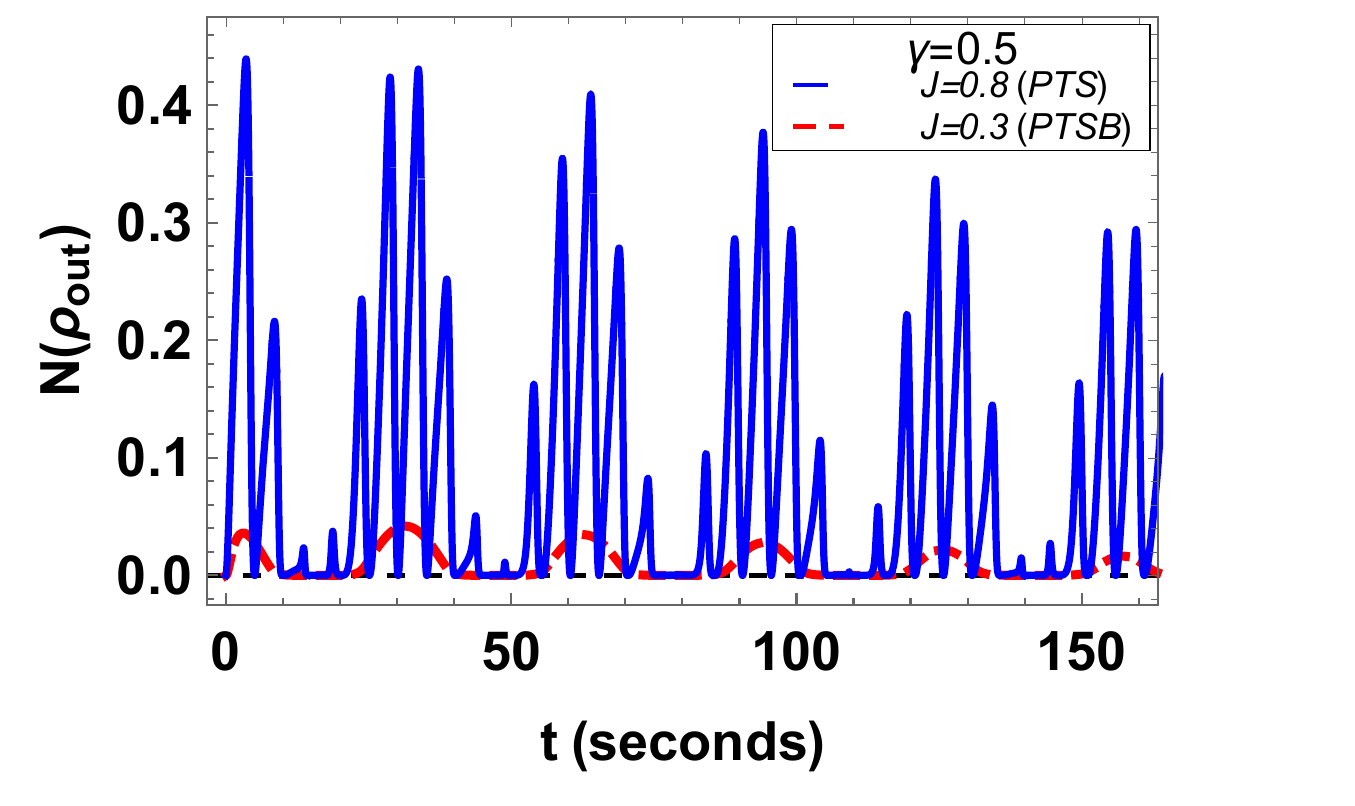}
	\includegraphics[width=65mm]{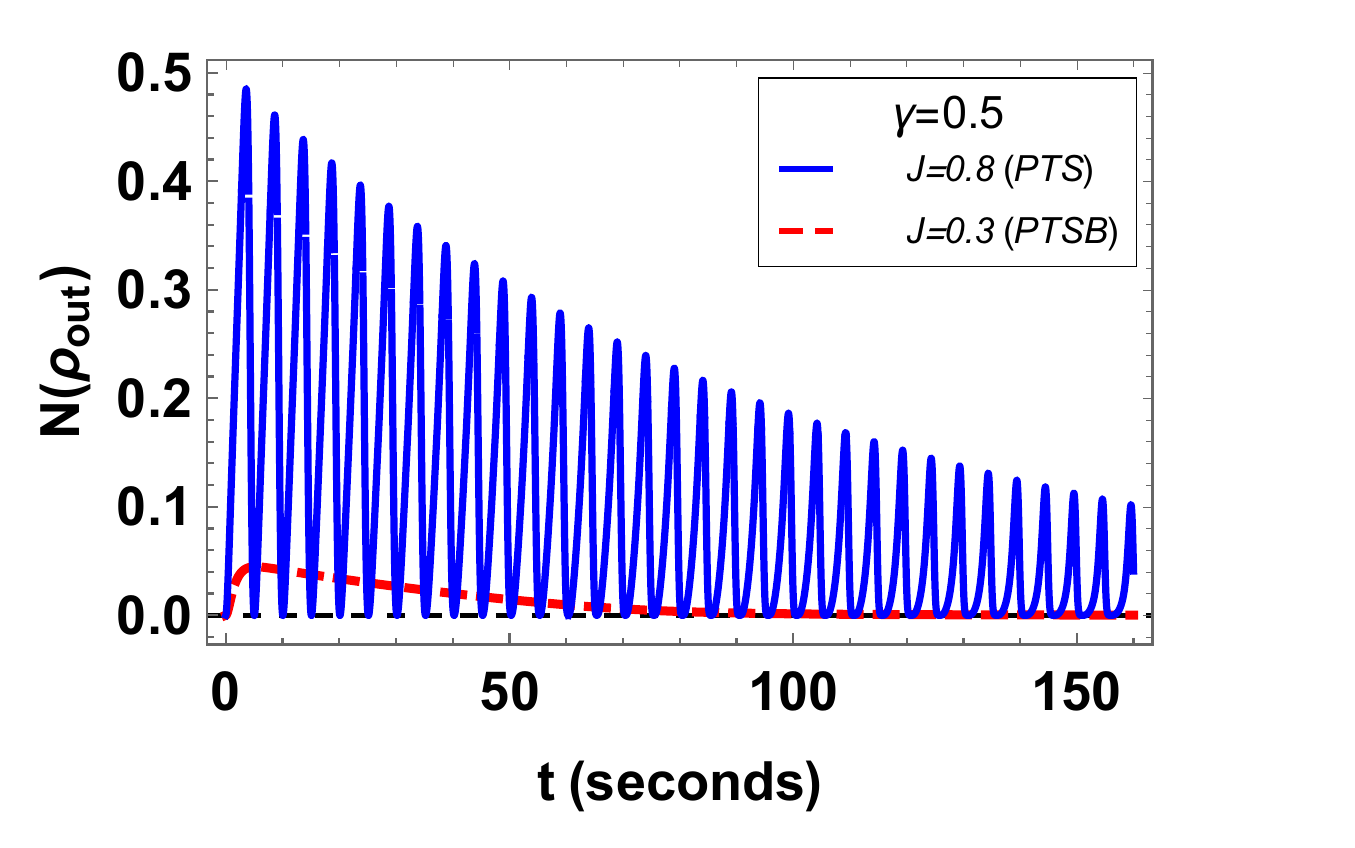}
	\caption{(Color online) Negativity  as a function of time $t$ when the output ports of the beam-splitter are subjected to RTN noise in non-Markovian (top) and  Markovian (bottom) regimes, respectively. The solid (blue) and dashed (red) curves correspond to PTS and PTSB phases, respectively.  The input state at the beam-splitter is given in Eq. (\ref{PTqubit}).}
	\label{fig:negativity}
\end{figure}

\section{Results and discussion}\label{ResultsDiscuss}
The  output state of the beam-splitter when the output ports are not subjected to noise channel is given in Eq. (\ref{rhoout}) which is in terms of the input state parameters $p$ and $x$. We consider the case when the input state is  given by Eq. (\ref{PTqubit}).  This allows us to study the interplay between nonclassiclaity of the state and the $\mathcal{PT}$ symmetry of the underlying system Fig. \ref{Model}. The real eigenvalues, Fig. \ref{fig:eigenvalues}, imply the complete $\mathcal{PT}$ symmetry  of the underlying effective Hamiltonian, Eq. (\ref{eq:Heff}).\par
In order to investigate the nonclassicality of the output state, we study the well known measures such as MID,  concurrence, and  negativity. These quantities are depicted in Fig. \ref{fig:MID_Con_Neg} with respect to time. Enhancement in the magnitude of these measures is observed in PTS phase compared to PTSB phase. This enhancement  in nonclassicality can be attributed to the fact that in $\mathcal{PT}$ symmetric phase, the coupling strengths dominate the loss/gain rate.\par
Things become interesting when the output ports of the beam-splitter are subjected to the noisy quantum channels as illustrated in Fig. \ref{fig:BSchannel}. In this work, we considered three important channels, viz., RTN, PD and AD. The channels could be same or different depending on the channel parameters. We have considered the same channels on both output ports. In Fig. \ref{conc_peq1}, the concurrence of the output state, when the input state is $(0~1)^T$, is depicted with respect to time. Under the RTN evolution,the dynamics in non-Markovian and Markovian regimes is contrasted by the characteristic recurrent behavior in the former case.  The concurrence evolves under the RTN channel as follows: 
\begin{align}
C(\rho_{out})|_{p=1} &= e^{(- \tilde{\gamma}_1 t) } \big[\cos( \mu_1  \tilde{\gamma}_1 t )  + \frac{\sin(\mu_1  \tilde{\gamma}_1 t )}{\mu_1}\big]\nonumber \\&\times e^{(- \tilde{\gamma}_2 t) } \big[\cos( \mu_2  \tilde{\gamma}_2 t )  + \frac{\sin(\mu_2  \tilde{\gamma}_2 t )}{\mu_2}\big]
\end{align}
 The evolution is non-Markovian or Markovian according as $\mu_i$ ($i=1,2$) is real or imaginary. We have considered the case in which the channels on the output ports are identical, that is, $\tilde{\gamma}_1 = \tilde{\gamma}_2$ and $\mu_1 = \mu_2$. In non-Markovian regime, one  observes an enhancement in the concurrence and thus the entanglement as compared with the Markovian case. The behavior of concurrence in PD and AD channels shows the typical decrease with time.\par
 We now analyze the behavior of various measures of nonclassicality when the input state is that of the effective two level atom given by Eq. (\ref{PTqubit}). The output state is subjected to various noise models. Figure (\ref{fig:MID}) shows the measurement induced disturbance (MID) quantified by the parameter $Q (\rho_{out})$.  Although, the application of the noise channels results in decreasing the degree of nonclassicality (quantified by the maximum value) of various measures, nevertheless, the nonclassicality measures continue to show distinct  behavior  in PTS and PTSB regimes, with enhanced magnitude in the former case.  Further, in case of RTN noise, the non-Markovian regime is seen to show the typical recurrent behavior. Similar behavior is observed in case of concurrence Fig. \ref{fig:concurrence} and negativity Fig.(\ref{fig:negativity}. The oscillating feature of the  measures of nonclassiclaity in the Markovian regime of RTN as well as in case of PD and AD channels should not be confused with the characteristic recurrent behavior of non-Markovian dynamics. This feature appears due to the oscillatory nature fo the input state given in Eq. (\ref{PTqubit}). This becomes clear when one looks at Fig. \ref{conc_peq1}, where the input state is $\ket{1} = (0~~1)^T$, $T$ being the transpose operation. 

\section{Conclusion}\label{conclusion}
We considered a $\Lambda$ type three level atom and derived an effective two level system bearing $\mathcal{PT}$ symmetry. The $\mathcal{PT}$ symmetry is governed by the coupling strength between the two levels and their respective loss/gain rate, such that when the coupling dominated the loss/gain rate, the system is in $\mathcal{PT}$ symmetric phase. However, when the gain/loss dominates the coupling, the system is said to be in $\mathcal{PT}$ broken phase. Consequently, the eigenvalues of the underlying effective Hamiltonian are real and imaginary in the former and later cases, respectively.\par
The beam-splitter provides an elegant way of analyzing the nonclassical properties of the output state when the qubit at the input is combined with  vacuum. We investigated various measures of nonclassicality, viz., measurement induced disturbance (MID), concurrence, and negativity of the output state when the input state is the effective qubit state of our $\mathcal{PT}$ symmetric system. The nonclassical measures behave quite different in  PTS and PTSB regimes, depicting an enhancement of nonclassicality in the former case.  Further, the application of various noise channels is accompanied with a decrease in the  degree of nonclassicality, however, the nonclassicality measures continue to show distinct behavior in PTS and PTSB phases, dominating in the former case.  We considered three noise channels viz., non-Markovian Random Telegraph Noise (RTN) as well as Markovian Phase Damping (PD), and Amplitude Damping (AD) channels and analyzed the behavior of various nonclassical measures under the influence of these channels. This study, therefore brings an interesting interplay of the quantumness of a system, along with its memory, and its $\mathcal{PT}$ symmetry, a property which is controlled by the coupling strength and the loss/gain rate associated with the energy levels of the system.
The conceptual ideas and methods introduced here are quite general, and the same can be used to study the dynamics of various other physical systems. For example, a comparative study of all types of three level systems \cite{Amarendra,LiOptExpr} in the context $\mathcal{PT}$ symmetry  is ongoing currently.
\FloatBarrier
\section*{Appendix}
In order to analyze the effect of various quantum channels on the nonclassical properties of the output state $\rho_{out}$ subjected to quantum noise channels, we rewrite the general bipartite state in a useful form. Consider a general two party system with the underlying Hilbert spaces denoted by $\mathcal{H}_1$ and $\mathcal{H}_2$ with the corresponding bases $\{ \ket{0}_1, \ket{1}_1\}$ and  $\{ \ket{0}_2, \ket{1}_2\}$, respectively. The tensor product state can be defined as
\begin{align}\label{eq:TensorP}
\ket{w} &=  \beta_{00} \ket{0}_1 \otimes \ket{0}_2  + \beta_{01} \ket{0}_1 \otimes \ket{1}_2 \\&+ \beta_{10} \ket{1}_1 \otimes \ket{0}_2 + \beta_{11} \ket{1}_1 \otimes \ket{1}_2.
\end{align}
This can also be viewed as a matrix 
\begin{equation}\label{Mw1}
M_w = (\beta_{ij}) = \begin{pmatrix}
a & b\\  
c & d
\end{pmatrix}.
\end{equation}
According to the \textit{singular value decomposition} theorem, any matrix can be written as a product $U \Sigma V^\dagger$, where both $U$ and $V$ are orthogonal matrices and $\Sigma$ is a diagonal matrix. The elements of $D$ are called \textit{singular values}. Therefore, we have 
\begin{equation}\label{SVD}
M_{w} =U \Sigma V^\dagger. 
\end{equation}
In order to diagonalize $M_w$, we note that the columns of both $U$ and $V$ matrices form orthogonal basis, therefore 
\begin{equation}
U = \begin{pmatrix}
\ket{u_0} & \ket{u_1} 
\end{pmatrix} \qquad {\rm and }~~~~V^\dagger =\begin{pmatrix}
\bra{v_0} \\  \bra{v_1}
\end{pmatrix}.
\end{equation}
Hence,
\begin{align}\label{Mw}
M_w = U D V^\dagger &= \begin{pmatrix}
\ket{u_0} & \ket{u_1} 
\end{pmatrix} \begin{pmatrix}
\sigma_+   & 0\\
0        & \sigma_-
\end{pmatrix}       \begin{pmatrix}
\bra{v_0} \\  \bra{v_1}
\end{pmatrix}.                                 
\end{align}
Here $\sigma_{\pm}$ are the \textit{singular values} (by definition, they are the elements of the diagonal matrix $\Sigma$, Eq. (\ref{SVD})). If one writes $M_w = z_0 \mathbf{I} + z_1 \sigma_1 + z_2 \sigma_2 + z_3 \sigma_3$, where $\sigma_i$ are Pauli matrices, Then the expression for $\sigma_{\pm}$ turns out to be 
\begin{equation}
\sigma_{\pm} = \sqrt{ \sum\limits_{i=0}^3 |z_i|^2 \pm \sqrt{\Big(\sum\limits_{i=0}^3 |z_i|^2 \Big)^2 - | z_0^2-z_1^2-z_2^2-z_3^2 |^2} }.
\end{equation}
Here, $z_0 = (a+d)/2$, $z_1 = (b+c)/2$, $z_2 = i(b-c)/2$ and $z_3 = (a-d)/2$, where $a,~b,~c$ and $d$ are as defined in Eq. (\ref{Mw1}). With this, the singular values are given by
\begin{equation}
\sigma_{\pm} = \sqrt{\frac{1}{2} \pm \sqrt{\frac{1}{4} - |ad -bd|^2}}.
\end{equation}
Here, use has been made of $|a|^2 + |b|^2 + |c|^2 + |d|^2 =1$. Since $\sigma_+^2 + \sigma_-^2 = 1$, we redefine $\sigma_+ = \sqrt{\alpha}$ and $\sigma_- = \sqrt{ 1 - \alpha}$. Therefore, from Eq. (\ref{Mw}), we have
\begin{equation}
M_w = \sqrt{\alpha} \ket{u_0} \bra{v_0} + \sqrt{1- \alpha} \ket{u_1}\bra{v_1}.
\end{equation}
Using this in Eq. (\ref{eq:TensorP}), we see that the tensor product state becomes
\begin{equation}
\ket{w} = \sqrt{\alpha} \ket{u_0} \ket{v_0} + \sqrt{1 - \alpha} \ket{u_1} \ket{v_1}.
\end{equation}
Setting $\ket{u_0} = \ket{0}_u$, $\ket{u_1} = \ket{1}_u$, $\ket{v_0} = \ket{1}_v$ and $\ket{v_1} = \ket{0}_v$, we have 
\begin{equation}
\ket{w} = \sqrt{\alpha} \ket{01}  + \sqrt{1 - \alpha} \ket{10}.
\end{equation}
This, however, should not be interpreted as the generation of an entangled state from a separable state. A separable state $\ket{\psi} = \frac{1}{2}(\ket{00} + \ket{01} + \ket{10} + \ket{11})$, with $a=b=c=d = 1/2$,  leads to $\alpha = 1$, and hence $\ket{w} = \ket{01}$.
%
\end{document}